\documentclass[10pt,conference]{IEEEtran}

\usepackage{cite}
\usepackage{amsmath,amssymb,amsfonts}
\usepackage{textcomp}
\usepackage{fancyhdr}

\usepackage{tikz}
\usepackage{placeins}
\usepackage{graphicx}
\usepackage{subcaption}
\usepackage[hyphens]{url}
\usepackage{color}
\usepackage[dvipsnames,svgnames,table]{xcolor}
\usepackage{multirow}
\usepackage{lipsum}
\usepackage{caption}
\usepackage{wrapfig}

\usepackage{makecell}
\usepackage{xspace}
\usepackage{pifont}

\usepackage[linecolor=DarkGreen,backgroundcolor=DarkGreen!25,bordercolor=DarkGreen]{todonotes}

\usepackage[noend, ruled, vlined, linesnumbered, noline, boxed]{algorithm2e}

\usepackage{booktabs}
\usepackage{etoolbox}

\usepackage{adjustbox}
\usepackage{bm}
\usepackage{soul}
\newcommand{\minisection}[1]{\vspace{0.5ex}\noindent {\bf #1}}

\definecolor{DarkBlue}{RGB}{0, 70, 140}
\definecolor{DarkRed}{RGB}{170, 0, 0}

\newcommand{\sym}[1]{\textsf{#1}}

\newcommand{\name}{\sym{Lottery BP}{}\xspace}
\newcommand{\namefull}{\sym{Lottery BP+OSD}{}\xspace}
\newcommand{\namearch}{\sym{PolyQec}{}\xspace}
\newcommand{\namesim}{\sym{Syndrilla}{}\xspace}

\newcommand{\thetitle}{\name: Unlocking Quantum Error Decoding at Scale}

\newcommand{\keywords}[1]{%
  \begin{IEEEkeywords}#1\end{IEEEkeywords}}

\usepackage[most]{tcolorbox}

\DeclareRobustCommand{\circled}[1]{%
  \tikz[baseline=(char.base)]{
    \node[shape=circle,draw=black,fill=black,inner sep=1pt, line width=0.75pt] (char)
    {\color{white}\sffamily\scriptsize #1};}}

\usepackage{threeparttable}
\usepackage{fontawesome}
\newcounter{insightcounter}
\newcommand{\insight}[2][]{
  \refstepcounter{insightcounter}
  \begin{tcolorbox}[
    colframe=gray,
    colback=lightgray,
    boxrule=0.5pt,
    arc=3pt,
    left=0pt,
    right=0pt,
    top=0pt,
    bottom=0pt
  ]
  \ifx\\#1\\%
    \textbf{\faLightbulbO\ Insight \arabic{insightcounter}}: #2
  \else
    \hypertarget{#1}{}%
    \label{#1}%
    \textbf{\faLightbulbO\ {Insight \ref*{#1}}}: #2
  \fi
  \end{tcolorbox}
}

\usepackage{hyperref}

\newcommand{\hpcayear}{2027}

\newcommand{\hpcasubmissionnumber}{2036}
\title{\thetitle}
\def\hpcacameraready{} % Uncomment to build camera-ready version

\newcommand\hpcaauthors{Yanzhang Zhu$^\dagger$, Chen-Yu Peng$^\ddagger$, Yun Hao Chen$^\ddagger$, Yeong-Luh Ueng$^\ddagger$, and Di Wu$^\dagger$}
\newcommand\hpcaaffiliation{$^\dagger$Department of ECE, University of Central Florida, Orlando, FL, USA \\
        $^\ddagger$Department of EE, National Tsing Hua University, Hsinchu, Taiwan}
\newcommand\hpcaemail{\{yanzhang.zhu, di.wu\}@ucf.edu, \{ss113061597, sdfg0424\}@gapp.nthu.edu.tw, ylueng@ee.nthu.edu.tw
}

\begin{document}
%% Begins the document, sets the (blind) author banner, \maketitle, page style.
%%%%%%%%%%%%%%%%%%%%%%%%%%%%%%%%%%%%%
%%%%%%%%%% -- DO NOT MODIFY -- %%%%%%%%%%
%%%%%%%%%%%%%%%%%%%%%%%%%%%%%%%%%%%%%

\author{
  \ifdefined\hpcacameraready
    \IEEEauthorblockN{\hpcaauthors{}}
      \IEEEauthorblockA{
        \hpcaaffiliation{} \\
        \hpcaemail{}
      }
  \else
    \IEEEauthorblockN{\normalsize{HPCA \hpcayear{} Submission
      \textbf{\#\hpcasubmissionnumber{}}} \\
      \IEEEauthorblockA{
        Confidential Draft \\
        Do NOT Distribute!!
      }
    }
  \fi 
}

% Heading and footer for title page
\fancypagestyle{camerareadyfirstpage}{%
  \fancyhead{}
  \renewcommand{\headrulewidth}{0pt}
  \fancyhead[C]{
    \ifdefined\aeopen
    \parbox[][12mm][t]{13.5cm}{\hpcayear{} IEEE International Symposium on High-Performance Computer Architecture (HPCA)}    
    \else
      \ifdefined\aereviewed
      \parbox[][12mm][t]{13.5cm}{\hpcayear{} IEEE International Symposium on High-Performance Computer Architecture (HPCA)}
      \else
      \ifdefined\aereproduced
      \parbox[][12mm][t]{13.5cm}{\hpcayear{} IEEE International Symposium on High-Performance Computer Architecture (HPCA)}
      \else
      \parbox[][0mm][t]{13.5cm}{\hpcayear{} IEEE International Symposium on High-Performance Computer Architecture (HPCA)}
    \fi 
    \fi 
    \fi 
    \ifdefined\aeopen 
      \includegraphics[width=12mm,height=12mm]{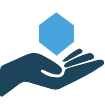}
    \fi 
    \ifdefined\aereviewed
      \includegraphics[width=12mm,height=12mm]{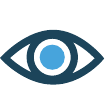}
    \fi 
    \ifdefined\aereproduced
      \includegraphics[width=12mm,height=12mm]{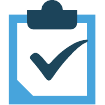}
    \fi
  }
  %\fancyfoot[L]{\hpcapubid{} \copyright \hpcayear{} IEEE}
  \fancyfoot[C]{}
}
% Heading and footer for remaining pages
\fancyhead{}
\renewcommand{\headrulewidth}{0pt}
%\fancyhead[C]{\hpcayear{} IEEE International Symposium on
% High-Performance Computer Architecture (HPCA)}

\maketitle

\thispagestyle{plain}
\pagestyle{plain}

%Enables the camera ready header and footer
% \ifdefined\hpcacameraready 
%   \thispagestyle{camerareadyfirstpage}
%   \pagestyle{empty}
% \else
%   \thispagestyle{plain}
%   \pagestyle{plain}
% \fi

% \newcommand{\hpcaheight}{0mm}
% \ifdefined\eaopen
% \renewcommand{\hpcaheight}{12mm}
% \fi

%%%%%%%%%%%%%%%%%%%%%%%%%%%%%%%%%%%%%%%%%%%%%%%%%%%%%%%%%%%%%%%%%%%%%%%%%%
%% Paper content.
%%%%%%%%%%%%%%%%%%%%%%%%%%%%%%%%%%%%%%%%%%%%%%%%%%%%%%%%%%%%%%%%%%%%%%%%%%

\begin{abstract}
Entering the megaquop era with millions of qubits towards quantum utility, quantum error correction (QEC) is leveraged to achieve fault tolerant quantum computing (FTQC).
QEC encodes multiple physical qubits into one logical qubit to protect quantum information from noises of varying forms.
During a QEC cycle, quantum error decoding stands on the critical path.
To enable fault tolerance on millions of qubits in real time, scalable decoding is necessary, which motivates this paper.
Existing decoding algorithms (decoders), such as clustering, matching, belief propagation (BP), and neural networks, suffer from one or more of inaccuracy, costliness, and incompatibility, upon a broad set of quantum error correction codes, such as surface code and bivariate bicycle code.
Therefore, there exists a gap between existing decoders and an ideal decoder that is accurate, fast, general, and scalable simultaneously.

% \todo{update numbers}
To move closer to the goal above, this paper contributes in three aspects, including decoder algorithm, decoder architecture, and decoding simulator.
\underline{First}, we propose \textit{\name}, a lightweight decoder that introduces guided randomness to break the symmetric deadlock caused by quantum degeneracy during decoding. 
% \todo{can expand a bit about the algorithm, what is the highlight of the algorithm?}
\name improves the decoding accuracy over BP by up to 6 orders.
% To efficiently decode multi-round measurement errors, we propose \textit{syndrome vote} as a pre-processing step before \name, that performs majority vote on multiple rounds of syndromes to compress them into one.
% Syndrome vote effectively increases the latency margin of decoding and mitigates the backlog problem.
\underline{Second}, we design a \textit{\namearch} architecture that implements \name as a local decoder and ordered statistics decoding (OSD) as a global decoder, exemplifying a hierarchical decoder architecture.
% \todo{is this still true? Yes, it accept both X/Z, the data need to updated once the invoke rate for Zhengyou has came out}
\namearch is configurable for surface code and X/Z check.
Since \name boosts the local decoding accuracy, \namearch invokes the costly global OSD decoder less frequently over BP+OSD to enhance the scalability, e.g., up to 4 orders of magnitude less for surface codes.
% \todo{Syndrilla statement is too long, make it shorter, three should be balanced}
\underline{Third}, we develop \textit{\namesim}, a modular PyTorch-based decoding simulator that enables fair, extensible decoder evaluation with unified accuracy and performance metrics.
On GPUs, \namesim runs $1$ order of magnitude faster than CUDAQX.
% \underline{Third}, to evaluate decoders fairly, we develop a PyTorch-based decoding simulator, \textit{\namesim}, that modularizes the simulation pipeline and allows to extend new decoders flexibly.
% We formulate multiple metrics to quantify the performance and accuracy of decoders and integrate them in \namesim.
% Running on GPUs, \namesim is $1\sim2$ orders of magnitude faster than CPUs with identical simulation fidelity across varying data formats.
\end{abstract}


\section{Introduction}
\label{sec:Introduction}

Quantum computing has been envisioned to revolutionize science and engineering, such as physics~\cite{di2024quantum, ayral2023quantum}, biomedicine~\cite{durant2024primer, chow2024quantum}, chemistry~\cite{cao2019quantum, lanyon2010towards}, and cryptography~\cite{fernandez2020towards, mavroeidis2018impact}, by offering exponential speedup in computation~\cite{jozsa2003role}.
Varying forms of noises have kept quantum computing in the NISQ era for years~\cite{preskill2018quantum}.
To seek quantum utility with millions of qubits in the megaquop era~\cite{herrmann2023quantum, preskill2025beyond}, fault-tolerant quantum computing is mandatory, especially via quantum error correction (QEC)~\cite{terhal2015quantum, cory1998experimental, devitt2013quantum}.
QEC encodes multiple physical qubits (data and syndrome qubits) into fewer logical qubits via QEC code.
Based on the threshold theorem~\cite{knill2005quantum, aliferis2005quantum}, the logical error rate can be exponentially suppressed with larger code distances.
By measuring the syndrome, decoders can infer which data qubits are erroneous with a certain accuracy.

\minisection{Quality factors.}
There exist multiple factors that impact the quality of QEC.
\underline{First}, decoders must be accurate enough to reach a high or even infinite circuit depth.
There have been many decoders proposed, such as tensor network~\cite{PhysRevLett.113.030501}, minimum-weight perfect-matching (MWPM)~\cite{MWPM_1, MWPM_2, MWPM_3}, union-find (UF)~\cite{union_find_paper}, neural networks~\cite{lange2025data, liu2019neural, nvidia_qec_decoder}, and belief propagation (BP)~\cite{syndrome_based_BP_paper_3, degeneracy_2}, with varying decoding accuracy.
Inaccurate decoders will interrupt the logical operations frequently, lowering performance.
\underline{Second}, decoders need to be fast enough, i.e., faster than syndrome generation (e.g., 400ns for superconducting~\cite{surface_cycle}), to mitigate the backlog problem~\cite{terhal2015quantum, holmes2020nisq+}.
Slow decoders fail to decode the syndrome timely and cause stalls.
Moreover, due to measurement errors, one QEC cycle must include multiple measurement rounds~\cite{MWPM_1} to form more complex 3D space-time decoding.
% By constructing a 3D decoding graph, measurement errors can be successfully identified, especially for matching-based decoders (e.g., UF and MWPM).
\underline{Third}, decoders need to be general enough to decode different QEC codes.
Different QEC codes, such as surface code~\cite{fowler2012surface}, quantum low-density parity-check (QLDPC) code~\cite{panteleev2021degenerate, BB_code_paper}, and color code~\cite{landahl2011fault}, have distinct code construction, that could be unfriendly to certain decoders.
For example, UF and MWPM fail to decode QLDPC code with high dimensionality and non-local connection, while BP fails to show a threshold~\cite{stephens2014fault, bposdcpp}.
\underline{Fourth}, decoders have to be scalable enough to support parallel decoding across millions of qubits~\cite{afs_paper}.
If the classical decoding system cannot scale up for massively parallel decoding, logical errors could accumulate and propagate, limiting the system scale.

% talk about scalability, reconfigurability for decoder system
% more...
\begin{figure}[!t]
    \centering
    \includegraphics[width=\linewidth]{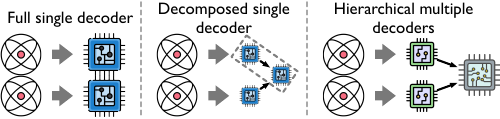}
    \caption{Design space of decoder architecture.
    The chip size represents the relative cost to decode a logical qubit, while the chip type denotes decoding algorithms.}
    \label{fig:design_space_overview}
\end{figure}

\minisection{Limitations.}
The design space of decoder architecture is shown in Figure~\ref{fig:design_space_overview}.
The \underline{left} architecture allocates one full decoder for one or more logical qubit, and the decoding cost is linear to the number of qubits~\cite{Vegapunk, Micro_blossom, astrea, das2022lilliput}.
This architecture usually requires the decoder is cheap enough to implement, e.g., via look-up table (LUT)~\cite{das2022lilliput} or approximation~\cite{astrea}.
This method is often limited to small code distances, since they tend to search the error space in a brute-force manner, which grows exponentially with the code distance.
Furthermore, these decoders are usually matching-based and not compatible to QLDPC code.
The \underline{middle} architecture decomposes a single decoder into multiple pipelined stages and allocates a different number of subcomponent for each stage, depending on how each stage is utilized during decoding~\cite{afs_paper}.
This organization scales better than a full single decoder per group qubits, and the gain stems from better utilizing the resource within a single decoder.
However, existing decoders like AFS~\cite{afs_paper} adopts a UF decoder and struggles with decoding accuracy~\cite{10.1145/3575693.3575733, astrea}.
The \underline{right} architecture combines two types of decoders to construct local-global, hierarchical decoding~\cite{10.1145/3575693.3575733, bposdcpp}, where the local decoder is way simpler than that of the global decoder.
One example, Clique~\cite{10.1145/3575693.3575733}, combines a local rule-based decoder in the superconducting domain and a global UF decoder in CMOS.
But the local decoder limits the overall scalability and accuracy~\cite{astrea}.
Another decoder, BP+OSD~\cite{bposdcpp}, combines local BP and global ordered statistics decoding (OSD).
% However, BP is known to be inaccurate for surface code due to quantum degeneracy~\cite{stephens2014fault, bposdcpp}, despite offering good accuracy for QLDPC code~\cite{yao2024belief}.
However, BP is known to be inaccurate for surface codes due to quantum degeneracy~\cite{stephens2014fault, bposdcpp}—despite offering good accuracy for QLDPC codes~\cite{yao2024belief}. This inaccuracy forces frequent invocations of the costly global OSD, completely neutralizing the latency benefits of the hierarchical design. 
% Moreover, \textit{the architecture exploration of BP+OSD has not yet been explored.}
Moreover, the hardware architecture of BP+OSD remains largely unexplored.

% \todo{update numbers}
\minisection{Proposal.}
In this paper, we aim to explore the algorithm and architecture design space of BP+OSD to enhance the accuracy, speed, scalability while maintaining generality to QEC codes.
\underline{First}, we introduce \name, a BP-based local decoder that introduces randomness during decoding to mitigate quantum degeneracy.
With low overhead, \name reduces the logical error rate by up to 6 orders of magnitude.
% We further propose syndrome vote as a pre-processing step for \name to compress multiple rounds of measured syndromes into one.
% Without losing accuracy, syndrome vote mitigates the backlog problem under measurement errors.
Combined with OSD, \namefull achieves comparable decoding accuracy as standard BP+OSD.
\underline{Second}, we design the \namearch architecture that implements \namefull for surface code and X/Z check.
We co-design the pipeline of \namearch to avoid throughput loss due to the algorithm innovation.
Due to the highly accurate local decoder, \namearch invokes the costly global OSD decoder by up to 4 orders of magnitude less frequently for better scalability for surface code.
\underline{Third}, we develop \namesim as a shared platform to fairly evaluate various decoders.
\namesim uses PyTorch to unlock hardware compatibility and GPU acceleration, and is modularized to enable extension to new decoders.
We also formulate multiple metrics to quantify the accuracy and performance of decoders and integrate them in \namesim.
Compared to CUDAQX~\cite{nvidia_qec_decoder} and StimBPOSD~\cite{stimbposd}, \namesim achieves over an order-of-magnitude higher simulation speed.
% \todo{need to double check whether algorithm and hardware architecture sections mention everything in the proposal paragraph. this should be done at proofreading phase.
% more accurate BP, backlog problem, low invoke => better scalability, general for topological and QLDPC, fast simulation}

\minisection{Contributions.}
The contributions of this work are as follows:

\begin{itemize}
    \item We innovate \name, a novel decoder to leverage randomness to mitigate quantum degeneracy errors. 
    \item We design \namearch architecture for \namefull to support surface code and X/Z check.
    % at different code distances.
    \item We develop \namesim, the first cross-platform framework to simulate quantum error decoding.
    \item We define novel metrics to quantify the accuracy and performance of decoders.
    \item We demonstrate the effectiveness of our proposed \name, \namearch, and \namesim.
\end{itemize}

In the following, Section~\ref{sec:Background} reviews the background.
Section~\ref{sec:Theory}, \ref{sec:Architecture}, and \ref{sec:Software} describe our \name algorithm, \namearch architecture, and \namesim simulator.
Section~\ref{sec:Implementation} and \ref{sec:Evaluation} evaluate our contributions.
Section~\ref{sec:Discussion} and \ref{sec:Conclusion} discuss related works and conclude this paper.

\section{Background}
\label{sec:Background}

\subsection{Decoding under Circuit-level noise}

\begin{figure}[!t]
    \centering
    \includegraphics[width=0.92\linewidth]{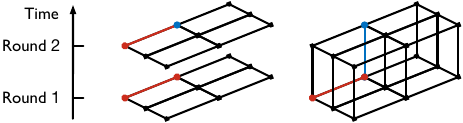}
    \caption{3D decoding graph for measurement error. Red and blue marks data qubit error and measurement error.}
    \label{fig:measurement_error_example}
\end{figure}
% \todo{This section needs to be replace by circuit level noise model?}

QEC codes protect quantum information by encoding physical qubits into logical ones. 
For instance, surface codes map qubits onto a 2D planar structure, naturally fitting superconducting topologies like IBM Nighthawk~\cite{mandelbaum2025qdc}.
However, realistic syndrome extraction suffers from \emph{circuit-level noise}~\cite{fowler2012surface}. 
Unlike 2D planar structure can only protect the logical qubits, circuit-level noise accounts for errors during state initialization, measurement, and multi-qubit entangling gates. 
These gate errors can propagate through the circuit, causing complex correlated faults on multiple qubits~\cite{gidney2021stim}.
% To tolerate this, one QEC cycle repeats measurements for $d$ consecutive rounds~\cite{MWPM_1} (Figure~\ref{fig:measurement_error_example}).
To capture complex gate-induced errors modern decoders (e.g., Micro Blossom~\cite{Micro_blossom}) rely on a Detector Error Model (DEM)~\cite{gidney2021stim}. 
The DEM maps independent circuit faults directly to flipped detectors, constructing a highly accurate graph that natively captures all circuit-level correlations.

% Multiple physical data and syndrome qubits are encoded into logical qubits via QEC codes to protect the quantum information.
% For example, surface code encodes physical qubits into 2D planar structure, which can be implemented easily on a superconducting quantum computer with 2D square-lattice topology, such as the recent IBM Nighthawk~\cite{mandelbaum2025qdc}.\
% However, realistic syndrome extraction suffers from \emph{circuit-level noise}~\cite{fowler2012surface}. 
% Unlike simplified phenomenological models that only consider independent data and readout faults, circuit-level noise accounts for errors during state initialization, measurement, and multi-qubit entangling gates. 
% These gate errors can propagate through the circuit, causing complex correlated faults on multiple qubits~\cite{gidney2021stim}.
% To tolerate such noise, one QEC cycle must include $d$ consecutive measurement rounds~\cite{MWPM_1} (Figure~\ref{fig:measurement_error_example}).
% A single measurement round (left) produces a conventional 2D syndrome, but its unreliability across rounds causes inconsistent decoding. 
% Instead, space-time decoding constructs a 3D decoding graph from these multiple rounds, where spatial and temporal edges capture data errors, measurement errors, and their circuit-level correlations.

\subsection{Decoding with BP and OSD}
% \todo{zhenyou, yunhao}
% \dw{talk about both single- and two-stage decoders}

\minisection{Belief propagation.} 
BP (or message passing) is a popular decoder to decode classical errors under low-density parity-check (LDPC) code~\cite{bp_paper,bp_ldpc}.
BP has nearly linear time complexity and high accuracy.
In QEC, hypergraph product (HGP)~\cite{hgp_paper}, one type of the Calderbank–Shor–Steane (CSS) code~\cite{css_paper}, can encode quantum information to two related LDPC codes.
Such code construction in HGP code allows to apply classical BP to decode QLDPC code~\cite{syndrome_based_BP_paper_1}.
% Since data qubits cannot be directly measured without collapsing the encoded information, decoding must rely on the measured syndrome from stabilizers \cite{Gottesman_stabilizer}.
Since QEC relies on the measured syndrome from stabilizers \cite{Gottesman_stabilizer}, a syndrome-based belief propagation framework is used \cite{syndrome_based_BP_paper_1, wang2012enhanced, syndrome_based_BP_paper_3}, differing from classical BP taking data as input.
Unlike matching/clustering-based decoders (e.g., UF and MWPM) that treat data and syndrome qubits as nodes and edges, BP considers data and syndrome qubits as variable nodes (VNs) and check nodes (CNs), building a Tanner graph, where messages are passed iteratively between them.
Over iterations, the decoder will converge to an estimation of data qubit errors.
BP has many variants to balance speed and accuracy, and normalized min sum (NMS) algorithm is one of the most popular variants~\cite{norm_min_sum_paper, min-sum_paper}, that achieves almost identical accuracy and low cost.

\SetKwComment{Comment}{}{}
\newcommand{\CommentStyle}[2]{\textcolor{#1}{/*#2*/}}
\begin{algorithm}[!t]
% \small
\caption{Normalized min sum BP algorithm~\cite{norm_min_sum_paper} in blue lines.
Our proposed \name additionally includes a light-weight lottery policy in the red.
\label{alg:decoding_nms_bp}}
\KwIn{
    Parity-check matrix $H$, 
    max decoding iteration $I$ (decided by code distance $d$), 
    physical error rate $p$, 
    measured syndrome $s$.
}
\KwOut{
    Estimated error $\hat{e}$.
}
$\displaystyle\mu_{v} = \ln{\frac{1-p_v}{p_v}}$ \Comment*[r]{\CommentStyle{DarkBlue}{Prior LLR}}
\For{$i = 0$ \KwTo $I-1$}{
    \eIf{$i==0$}{
        $\displaystyle\alpha^{(i)}_{v\rightarrow c} = \mu_v$
        \Comment*[r]{\CommentStyle{DarkBlue}{Initialization}}
    }{
        % VN update
        $\displaystyle\alpha^{(i)}_{v\rightarrow c} = \lambda^{(i-1)}_v - \beta^{(i-1)}_{v\leftarrow c}$
        \Comment*[r]{\CommentStyle{DarkBlue}{VN update}}
    }

    % \hrulefill
    
    $\displaystyle\beta^{(i)}_{v\leftarrow c} = (-1)^{s_c} \cdot \left( 1-\frac{1}{2^{i+1}} \right) \cdot \prod_{v^{\prime} \in \mathcal{V}(c) \setminus \{v\}} \operatorname{sign}\left(\alpha^{(i)}_{v^{\prime}\rightarrow c}\right) \cdot \min_{v^{\prime} \in \mathcal{V}(c) \setminus \{v\}} |\alpha^{(i)}_{v^{\prime}\rightarrow c}|$
    \Comment*[r]{\CommentStyle{DarkBlue}{CN update}}
    % \hrulefill
    
    % LLR update
    $\displaystyle\lambda^{(i)}_{v} = \mu_{v} + \sum_{c \in \mathcal{C}(v)} \beta_{v\leftarrow c}^{(i)}$ \Comment*[r]{\CommentStyle{DarkBlue}{LLR update}}
    
    % Hard decision
    $\hat{e}^{(i)}_{v}=\textbf{Boolean}(\lambda^{(i)}_{v}\leq0)$\Comment*[r]{\CommentStyle{DarkBlue}{Hard decision}}

    % Syndrome estimation
    $\hat{s}^{(i)} = \hat{e}^{(i)}\cdot H^{\text{T}}\bmod 2$
    \Comment*[r]{\CommentStyle{DarkBlue}{Syndrome estimation}}

    % Termination check
    \If{$\hat{s}^{(i)} == s$}{
        % \hspace*{-0.5cm}\makebox[0.85\linewidth][l]{\hspace*{0.5cm}\Return $\hat{e}^{(i)}$\Comment*[r]{\CommentStyle{DarkBlue}{Early termination}}}
        \hspace*{-0.5cm}\underline{\makebox[0.85\linewidth][l]{\hspace*{0.5cm}\Return $\hat{e}^{(i)}$\Comment*[r]{\CommentStyle{DarkBlue}{Early termination}}}}
    }
    \underline{\makebox[0.85\linewidth][l]{\textbf{Lottery policy (Section~\ref{sec:lotterybp})}\Comment*[r]{\CommentStyle{DarkRed}{Lottery}}}}
}
\Return $\hat{e}^{(I-1)}$
\Comment*[r]{\CommentStyle{DarkBlue}{Max iteration termination}}
\end{algorithm}

We list NMS BP in Algorithm~\ref{alg:decoding_nms_bp}, which iterates three main steps until it converges, including VN update, CN update and Log-Likelihood Ratio (LLR) update. 
The full decoding procedure can be summarized as follows:
% \todo{double check the algorithm is correct, for example the entire line 6/7, including scaling factor, zhenyou/yanzhang}
\begin{itemize}
    \item \textbf{Prior LLR (line~1):}
    Since the intrinsic channel information is not available in quantum computing, the LLR for each VN $v$ is initialized using the uniform $p_v$, which is the model-dependent error probability of qubit $v$.
    % , which varies cross error models.
    
    \item \textbf{Initialization (line~4):}  
    At decoding iteration $i=0$, for each VN $v$ and its neighbor CN $c$, the variable-to-check (V2C) messages $\alpha^{(0)}_{v \to c}$ are initialized to the prior LLR.

    \item \textbf{Variable node update (line~6):}
    For $i > 0$, for each VN $v$ and its neighbor CN $c$, update the V2C message $\alpha^{(i)}_{v \to c}$ by subtracting the check-to-variable (C2V) message $\beta^{(i-1)}_{v \leftarrow c}$ from the posterior LLR $\lambda^{(i-1)}_v$ from the previous iteration.

    \item \textbf{Check node update (line~7):}  
    The C2V message $\beta^{(i)}_{v \leftarrow c}$ is computed by aggregating the V2C messages from all other neighboring VNs, i.e., $v^{\prime} \in \mathcal{V}(c) \setminus \{v\}$.
    Here, $s_c$ denotes the $c$-th component of the measured syndrome $s$, and $1-2^{-(i+1)}$ acts as a reliability scaling factor~\cite{min-sum_paper}. 
    % that adjusts the confidence level of the message.
    
    \item \textbf{LLR update (line~8):}  
    For each VN $v$, the posterior LLR $\lambda^{(i)}_v$ is updated by summing the prior LLR $\mu_v$ and all incoming C2V messages from the neighboring CNs.

    \item \textbf{Hard decision (line~9):}
    For each VN $v$, the estimated error $\hat{e}^{(i)}_v$ is 0 (no error) for $\lambda^{(i)}_v > 0$, and vice versa.
    
    \item \textbf{Syndrome estimation (line~10):}
    The estimated syndrome $\hat{s}^{(i)}$ is obtained via multiplication over the binary field $\mathbb{F}_2$.

    \item \textbf{Early termination (line~12):}
    If \( \hat{s}^{(i)} \) matches syndrome \( s \), decoding succeeds;
    otherwise, the iteration index $i$ increases by 1 and decoding repeats from line~2.
    
    \item \textbf{Max iteration termination (line~14):}
    If the estimated syndrome still does not match the measured syndrome, decoding is declared failed at the maximum iteration $I$.
\end{itemize}

%   \item Variable node (VN) update (Alpha update): Update all the alpha values that are stored in edges by checking the lambda value in each variable node they are connected to.
%   \item Check node update (Beta update): Update all the Beta values stored in edges by going through all the variable nodes connected to each check node.
%   \item Log-likelihood ratio update (LLR update): Update the lambda value in each variable node by going through all the check nodes that are connected to this variable node.
% \end{itemize}

BP can be used not only as a single-stage decoder, but also as an information extractor that produces soft or reliability information.
% about qubits.
The output can then serve as the input to a second-stage decoder, which significantly improves decoding accuracy~\cite{SSF_paper, syndrome_based_BP_paper_1, BP_GNN_paper, trapping_set_paper, gradient_descend_paper}.
% Our work expands to 
% Several algorithms have been proposed to leverage BP in this manner. 

% For instance, soft information from BP can guide the flipping decisions of small-set-flip (SSF), while its partial corrections can also help reduce the syndrome weight before SSF completes decoding \cite{SSF_paper}. This complementary interaction helps overcome the limitations of BP decoding in challenging cases \cite{syndrome_based_BP_paper_1} and mitigates SSF’s limited performance in high-weight syndrome cases.Numerical results demonstrate a notable improvement in decoding thresholds under noiseless syndrome conditions, along with competitive performance in the presence of noise.

% Another line of work integrates BP with Graph Neural Networks (GNNs) \cite{BP_GNN_paper}. Here, the output messages from BP are used as input features to intermediate GNN layers, which are interleaved between BP iterations. The GNN refines the messages and helps the decoder escape local minima caused by trapping sets or short cycles \cite{trapping_set_paper}. This integration significantly lowers the error floor compared to traditional BP-based methods.
% Although the GNN introduces some computational overhead, it leverages parallelism to improve computational efficiency and supports end-to-end training via gradient descent \cite{gradient_descend_paper}, which enhances decoding accuracy and robustness.
\begin{algorithm}[!t]
% \small
\caption{OSD-0 algorithm.
\label{alg:decoding_osd0}}
\KwIn{
    Parity-check matrix $H \in \mathbb{F}_2^{m \times n}$, 
    BP output LLRs $\lambda$, 
    measured syndrome $s \in \mathbb{F}_2^m$. 
}
\KwOut{
    Estimated error $\hat{e}_{\text{OSD-0}} \in \mathbb{F}_2^n$.
}

% Reliability sorting
$\pi \leftarrow \textbf{Sort}(\lambda)$ such that $\lambda_{\pi(1)} \leq \lambda_{\pi(2)} \leq \cdots \leq \lambda_{\pi(n)}$ 
\Comment*[r]{\CommentStyle{DarkBlue}{LLR sorting}}

% Column Permutation
$H_{\text{OSD}} \leftarrow H \cdot P_\pi$ 
\Comment*[r]{\CommentStyle{DarkBlue}{Column permutation}}

% Basis Selection
$S, H'_{\text{OSD}} \leftarrow$ First $\mathrm{rank}(H)$ independent columns of $H_{\text{OSD}}$ 
\Comment*[r]{\CommentStyle{DarkBlue}{Basis selection}}

% LU decomposition
$L, U \leftarrow \textbf{LU\_decompose}(H'_{\text{OSD}})$ 
\Comment*[r]{\CommentStyle{DarkBlue}{LU decomposition}}

% Forward substitution
$y \leftarrow \text{Solve } (L \cdot y = s)$ 
\Comment*[r]{\CommentStyle{DarkBlue}{Forward substitution}}

% Backward substitution
$e_S \leftarrow \text{Solve } (U \cdot e_S = y)$ 
\Comment*[r]{\CommentStyle{DarkBlue}{Backward substitution}}

% Error Vector Construction
$e' \leftarrow (e_S, \mathbf{0})$ 
\Comment*[r]{\CommentStyle{DarkBlue}{Error vector construction}}

% Inverse Permutation
$\hat{e}_{\text{OSD-0}} \leftarrow e'$ mapped back by $\pi^{-1}$ 
\Comment*[r]{\CommentStyle{DarkBlue}{Inverse mapping}}

\Return $\hat{e}_{\text{OSD-0}}$
\end{algorithm}

\minisection{Ordered-Statistics Decoding.}
Motivated by OSD for classical error correction~\cite{classical_osd_paper}, quantum versions of OSD, along with their higher-order variants, have been proposed for QLDPC and topological codes~\cite{bp_osd, bposdcpp}, by serving as the second stage in a BP+OSD decoder. 
When BP fails to produce a valid correction, OSD is invoked to estimate errors using the soft reliability from BP.
% Specifically, OSD constructs an information set based on this soft data and systematically explores low-weight error vector consistent with the measured syndrome.
This two-stage decoding combines the speed and efficiency of BP with the accuracy of OSD, lowering the error floor and improving the decoding speed.
Algorithm~\ref{alg:decoding_osd0} outlines OSD-0, the basis for high-order but more costly OSD variants:

\begin{itemize}
    \item \textbf{LLR sorting (line~1):}
    The output LLRs from BP are sorted in ascending order (from most likely to be an error to least likely), defining a permutation $\pi$.

    \item \textbf{Column permutation (line~2):} 
    The columns of matrix $H$ are reordered according to $\pi$ to form $H_{\text{OSD}}$.

    \item \textbf{Basis selection (line~3):} 
    Select the first $\text{rank}(H)$ linearly independent columns from $H_{\text{OSD}}$ to form the basis set $S$, and the corresponding basis submatrix $H'_{\text{OSD}}$.

    \item \textbf{LU decomposition (line~4):}
    Decompose $H'_{\text{OSD}}$ into a lower and upper triangular matrix $L$ and $U$ over $\mathbb{F}_2$, such that $H'_{\text{OSD}} \cdot e_S = s$ can be rewritten as $L \cdot U \cdot e_S = s$.

    \item \textbf{Forward substitution (line~5):}
    % The syndrome equation $H'_{\text{OSD}} \cdot e_S = s$ can now be rewritten as $L \cdot U \cdot e_S = s$. 
    Solve the lower triangular system for an intermediate vector $y$.
    
    \item \textbf{Backward substitution (line~6):}
    Solve the upper triangular system to find the estimated error vector $e_S$.
    
    \item \textbf{Error vector construction (line~7):}
    Concatenate $e_S$ with $\mathbf{0}$ as the error vector $e'$ in the permuted domain.
    % as, where $\mathbf{0}$ denotes an all-zero vector for the non-basis qubits.

    \item \textbf{Inverse mapping  (line~8):}
    Map $e'$ back to the original order as $\hat{e}_{\text{OSD-0}}$, which satisfies $H \cdot \hat{e}_{\text{OSD-0}} = s$.
\end{itemize}
% The OSD-0 procedure yields a valid error vector that satisfies the syndrome equation.
% Higher-order OSD variants improve accuracy over OSD-0, at much higher cost.
% has significantly lower computational complexity, which reduces decoding latency and makes it particularly attractive for hardware implementation.

\subsection{Challenges of BP for QEC}
\minisection{Quantum Degeneracy.}
% \todo{zhenyou, yunhao}
% A unique challenge in QEC that does not arise in classical error decoding is \emph{quantum degeneracy}~\cite{Gottesman_stabilizer}. 
% Two errors, $E_a$ and $E_b$, are considered to be degenerate if their product $E_a E_b$ lies in the stabilizer group $S$. 
Two errors are considered to be degenerate if their product lies in the stabilizer group~\cite{Gottesman_stabilizer}, which will further produce identical syndromes~\cite{Stabilizer_coset} that are indistinguishable to syndrome-based decoders, including BP.
% Since all errors within the same stabilizer coset produce identical syndrome patterns~\cite{Stabilizer_coset}, they are indistinguishable to any syndrome-based decoder, including BP.
% , as the measurement outcomes of the stabilizers are the only available input for inferring the error.
A large number of low-weight stabilizer generators increase the number of logically equivalent low-weight errors within stabilizer cosets~\cite{degeneracy, degeneracy_2}, and amplify the level of quantum degeneracy. 
This is particularly the case for surface codes at small code distances~\cite{syndrome_based_BP_paper_1, bp_degradation_2}, causing frequent decoding failures for BP.

\minisection{Short Cycles in Decoding Graph.}
The stabilizer code’s Tanner graph typically contains numerous short cycles~\cite{short_cycle}, which can introduce dependencies between messages.
Such dependencies could undermine the effectiveness of message passing and potentially hindering the convergence of BP~\cite{syndrome_based_BP_paper_1, bp_degradation_2, wang2012enhanced, babar2015fifteen}.

\minisection{Lacking Channel LLR.}
% Another challenge results from the lack of initial information of qubits in quantum decoding, 
% Unlike classical decoding, quantum decoding does not derive channel LLR from direct data qubit measurement~\cite{Gottesman_stabilizer}, but instead assumes identical channel LLR for all data qubits. 
Quantum decoding assumes identical prior LLR for all data qubits, due to lacking actual measurement~\cite{Gottesman_stabilizer}.
Such symmetry is problematic in the presence of degeneracy. 
% When CNs pass messages based on the measured syndrome, it is hard to distinguish which degenerate error patterns cause the unsatisfied syndrome. 
C2V messages for each error pattern are highly symmetric (i.e., split of belief~\cite{split_belief}), preventing a single error pattern from accumulating probability. 
Consequently, the BP decoder fails to distinguish which degenerate error patterns actually cause the unsatisfied syndrome. 
% This split of belief \cite{split_belief} distributes the posterior probabilities across multiple indistinguishable patterns, preventing any single error pattern from accumulating dominant probability. 
% Consequently, the BP decoder cannot converge to a unique solution that satisfies the measured syndrome.
% \todo{zhenyou, double check whether logic here is correct}

While optimal decoding is NP-hard~\cite{decoding_NP_hard}, this work introduces randomness in BP to mitigate quantum degeneracy. 

\subsection{Sign-Flipping-Related BP for QEC}

Sign-flipping BP perturbs the message passing to attack degeneracy directly: at selected iterations, the sign of a chosen VN's posterior LLR is flipped, breaking the belief symmetry that traps BP in a degenerate deadlock.
Prior methods differ in which VN and iteration to perturb, and each trades accuracy against latency differently.

\minisection{Branch-assisted sign-flipping BP (BSFBP).}
BSFBP~\cite{BSFBP} applies sign-flipping within a branched decoding process: when BP fails to converge, it spawns decoding branches, each flipping selected LLR signs of to explore a different perturbed trajectory. 
% \todo{we have not talk about it yet, no need to make it this concrete. just say selected from table xxx in sec xxx}
The LLR are chosen by one of several baseline selection policies (Table~\ref{tab:sf_policies} in Section~\ref{sec:lottery_design_space}).
This restores accuracy on degenerate cases, but the branching demands a large iteration budget. 
% orders of magnitude more iterations than plain BP, which inflates worst-case decoding latency.

\minisection{BP Guided decimation (BPGD).}
Rather than flipping a sign, BPGD~\cite{BPGD} freezes the LLR: after every $T$ iterations it selects the most reliable VN, fixes its LLR value, and restarts BP, achieving the final results VN-by-VN.
% It therefore acts globally and in the opposite direction of sign flipping, hardening the most reliable VN instead of perturbing the least reliable one.

\minisection{BP-Syndrome Flip (BP-SF).}
% \todo{why using this name? why not directly BP-SF, is there a name in paper?}
A third approach maintains a per-VN flip counter, incrementing it at iterations when the VN's decoded bit changes~\cite{BP_SF}.
% \todo{flip VN's what? sign}
If the BP iterations complete but not converge, it selects the VNs with the largest accumulated count and flips the syndrome bit for that VN, then reruns BP on the modified syndrome.

Across these methods, the perturbation is either costly in iterations (BSFBP), VN steps (BPGD), or costly in parallel hardware (BP-SF). 
% \todo{we can just say "we will study the design space of sign flip with the aim to lower the hardware cost."}
% None couples a hardware-friendly local scope with reliability-guided selection, the gap that \name (Section~\ref{sec:lotterybp}) targets.
Motivated by these trade-offs, we will study the design space of sign-flipping to lower its hardware cost without accuracy drop.
% which \name{} (Section~\ref{sec:lotterybp}) explores through a hardware-cheap local perturbation guided by BSFBP.

\section{Proposed \name Decoder}
\label{sec:Theory}

% This section will go over the intuition and design of \name, as well as syndrome vote as a pre-processing step for \name.

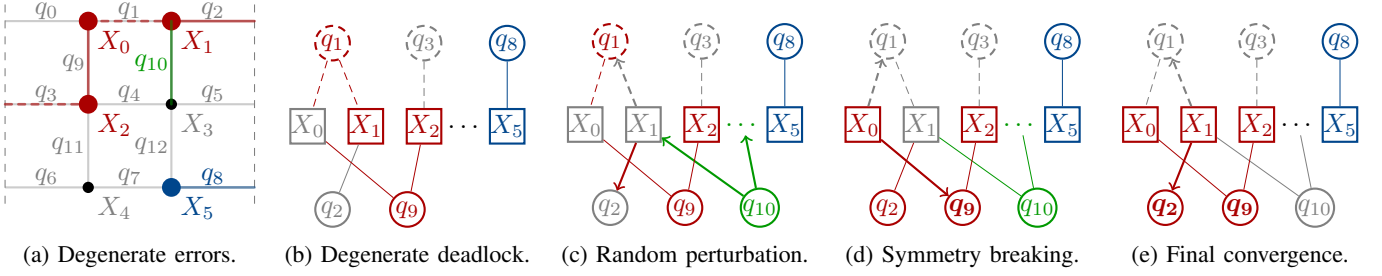
\begin{figure*}[!htbp]
    \centering
    
    \begin{subfigure}[b]{0.19\linewidth}
    \centering
    \begin{tikzpicture}[scale=1.1]
      \draw[thick, gray!40] (0,2) -- (1,2) node[midway, inner sep=0.5pt, text=gray, anchor=south] {$q_{0}$};
      \draw[thick, gray!40, anchor=south] (1,2) -- (2,2) node[midway, inner sep=0.5pt, text=gray] {$q_{1}$};
      \draw[thick, gray!40, anchor=south] (2,2) -- (3,2) node[midway, inner sep=0.5pt, text=gray] {$q_{2}$};
      \draw[thick, gray!40, anchor=south] (0,1) -- (1,1) node[midway, inner sep=0.5pt, text=gray] {$q_{3}$};
      \draw[thick, gray!40, anchor=south] (1,1) -- (2,1) node[midway, inner sep=0.5pt, text=gray] {$q_{4}$};
      \draw[thick, gray!40, anchor=south] (2,1) -- (3,1) node[midway, inner sep=0.5pt, text=gray] {$q_{5}$};
      \draw[thick, gray!40, anchor=south] (0,0) -- (1,0) node[midway, inner sep=0.5pt, text=gray] {$q_{6}$};
      \draw[thick, gray!40, anchor=south] (1,0) -- (2,0) node[midway, inner sep=0.5pt, text=gray] {$q_{7}$};
      \draw[thick, gray!40, anchor=south] (2,0) -- (3,0) node[midway, inner sep=0.5pt, text=DarkBlue] {$q_{8}$};

      \draw[thick, gray!40] (1,2) -- (1,1) node[midway, inner sep=0.5pt, text=gray, anchor=east] {$q_{9}$};
      \draw[thick, gray] (2,2) -- (2,1) node[midway, inner sep=0.5pt, text=green!60!black, anchor=east] {$q_{10}$};
      \draw[thick, gray!40] (1,1) -- (1,0) node[midway, inner sep=0.5pt, text=gray, anchor=east] {$q_{11}$};
      \draw[thick, gray!40] (2,1) -- (2,0) node[midway, inner sep=0.5pt, text=gray, anchor=east] {$q_{12}$};

      \draw[dashed, gray] (0, 2.2) -- (0, -0.2);
      \draw[dashed, gray] (3, 2.2) -- (3, -0.2);

      \node[circle, fill=black, inner sep=1.5pt] at (2,1) {}; 
      \node[below right, text=gray] at (2,1) {$X_3$};
      \node[circle, fill=black, inner sep=1.5pt] at (1,0) {}; 
      \node[below right, text=gray] at (1,0) {$X_4$};

      \node[circle, fill=DarkRed, inner sep=2.5pt] (X0) at (1,2) {}; 
      \node[below right, text=DarkRed] at (1,2) {$X_0$};
      \node[circle, fill=DarkRed, inner sep=2.5pt] (X1) at (2,2) {}; 
      \node[below right, text=DarkRed] at (2,2) {$X_1$};
      \node[circle, fill=DarkRed, inner sep=2.5pt] (X2) at (1,1) {}; 
      \node[below right, text=DarkRed] at (1,1) {$X_2$};
      \node[circle, fill=DarkBlue, inner sep=2.5pt] (X5) at (2,0) {}; 
      \node[below right, text=DarkBlue] at (2,0) {$X_5$};

      \draw[line width=1pt, DarkRed, opacity=0.6, line cap=round, dash pattern=on 2.5pt off 2.5pt] (1,2) -- (2,2);
      \draw[line width=1pt, DarkRed, opacity=0.6, line cap=round, dash pattern=on 2.5pt off 2.5pt] (0,1) -- (1,1);
      \draw[line width=1pt, DarkRed, opacity=0.6, line cap=round] (2,2) -- (3,2);
      \draw[line width=1pt, DarkRed, opacity=0.6, line cap=round] (1,2) -- (1,1);
      \draw[line width=1pt, DarkBlue, opacity=0.5, line cap=round] (2,0) -- (3,0);
      \draw[line width=1pt, green!60!black, opacity=0.5, line cap=round] (2,2) -- (2,1);
    \end{tikzpicture}
    \caption{Degenerate errors.}
    \label{fig:error_types}
    \end{subfigure}
    \hfill
    % --- (b) Step 1 ---
    \begin{subfigure}[b]{0.19\linewidth}
    \centering
    \begin{tikzpicture}[
        scale=1.1, 
        vn/.style={circle, draw=black, thick, minimum size=13pt, inner sep=0pt},
        cn/.style={rectangle, draw=black, thick, minimum size=13pt, inner sep=0pt}
    ]
    
      \node[cn, draw=gray, text=gray] (X0) at (0, 1) {$X_0$};
      \node[cn, draw=DarkRed, text=DarkRed] (X1) at (0.7, 1) {$X_1$};
      \node[cn, draw=DarkRed, text=DarkRed] (X2) at (1.4, 1) {$X_2$};
      \node[draw=none] (dots) at (1.9, 1) {$\dots$}; 
      \node[cn, draw=DarkBlue, text=DarkBlue] (X5) at (2.4, 1) {$X_5$};

      \node[vn, draw=DarkRed, text=DarkRed, densely dashed] (q1) at (0.3, 2) {$q_1$};
      \node[vn, draw=gray, text=gray, densely dashed] (q3) at (1.4, 2) {$q_3$};
      \node[vn, draw=DarkBlue, text=DarkBlue] (q8) at (2.4, 2) {$q_8$};

      \node[vn, draw=gray, text=gray] (q2) at (0.3, 0) {$q_2$};
      \node[vn, draw=DarkRed, text=DarkRed] (q9) at (1.2, 0) {$q_9$};

      \draw[very thin, DarkRed, densely dashed] (q1) -- (X0);
      \draw[very thin, DarkRed, densely dashed] (q1) -- (X1);
      \draw[very thin, gray, densely dashed] (q3) -- (X2);
      \draw[very thin, DarkBlue] (q8) -- (X5);
      \draw[very thin, gray] (q2) -- (X1);
      \draw[very thin, DarkRed] (q9) -- (X0);
      \draw[very thin, DarkRed] (q9) -- (X2);
    \end{tikzpicture}
    \caption{Degenerate deadlock.}
    \label{fig:step1}
    \end{subfigure}
    \hfill
    % --- (c) Step 2 ---
    \begin{subfigure}[b]{0.19\linewidth}
    \centering
    \begin{tikzpicture}[
        scale=1.1,
        vn/.style={circle, draw=black, thick, minimum size=13pt, inner sep=0pt},
        cn/.style={rectangle, draw=black, thick, minimum size=13pt, inner sep=0pt}
    ]
      \node[cn, draw=gray, text=gray] (X0) at (0, 1) {$X_0$};
      \node[cn, draw=gray, text=gray] (X1) at (0.7, 1) {$X_1$};
      \node[cn, draw=DarkRed, text=DarkRed] (X2) at (1.4, 1) {$X_2$};
      \node[draw=none, text=green!60!black] (dots) at (1.9, 1) {$\dots$};
      \node[cn, draw=DarkBlue, text=DarkBlue] (X5) at (2.4, 1) {$X_5$};

      \node[vn, draw=DarkRed, text=DarkRed, densely dashed] (q1) at (0.3, 2) {$q_1$};
      \node[vn, draw=gray, text=gray, densely dashed] (q3) at (1.4, 2) {$q_3$};
      \node[vn, draw=DarkBlue, text=DarkBlue] (q8) at (2.4, 2) {$q_8$};

      \node[vn, draw=gray, text=gray] (q2) at (0.3, 0) {$q_2$};
      \node[vn, draw=DarkRed, text=DarkRed] (q9) at (1.2, 0) {$q_9$};
      \node[vn, draw=green!60!black, text=green!60!black] (q10) at (2.1, 0) {$q_{10}$};

      \draw[very thin, DarkRed, densely dashed] (q1) -- (X0);
      \draw[->,  thick, gray, densely dashed] (X1) -- (q1);
      \draw[very thin, gray, densely dashed] (q3) -- (X2);
      \draw[very thin, DarkBlue] (q8) -- (X5);
      \draw[->,  thick, DarkRed] (X1) -- (q2);
      \draw[->,  thick, green!60!black] (q10) -- (X1);
      \draw[->,  thick, green!60!black] (q10) -- (dots);
      \draw[very thin, DarkRed] (q9) -- (X0);
      \draw[very thin, DarkRed] (q9) -- (X2);
    \end{tikzpicture}
    \caption{Random perturbation.}
    \label{fig:step2}
    \end{subfigure}
    \hfill
    % --- (d) Step 3  ---
    \begin{subfigure}[b]{0.19\linewidth}
    \centering
    \begin{tikzpicture}[
        scale=1.1,
        vn/.style={circle, draw=black, thick, minimum size=13pt, inner sep=0pt},
        cn/.style={rectangle, draw=black, thick, minimum size=13pt, inner sep=0pt}
    ]
      \node[cn, draw=DarkRed, text=DarkRed] (X0) at (0, 1) {$X_0$};
      \node[cn, draw=gray, text=gray] (X1) at (0.7, 1) {$X_1$};
      \node[cn, draw=DarkRed, text=DarkRed] (X2) at (1.4, 1) {$X_2$};
      \node[draw=none, text=green!60!black] (dots) at (1.9, 1) {$\dots$};
      \node[cn, draw=DarkBlue, text=DarkBlue] (X5) at (2.4, 1) {$X_5$};

      \node[vn, draw=gray, text=gray, densely dashed] (q1) at (0.3, 2) {$q_1$};
      \node[vn, draw=gray, text=gray, densely dashed] (q3) at (1.4, 2) {$q_3$};
      \node[vn, draw=DarkBlue, text=DarkBlue] (q8) at (2.4, 2) {$q_8$};

      \node[vn, draw=DarkRed, text=DarkRed] (q2) at (0.3, 0) {$q_2$};
      \node[vn, draw=DarkRed, text=DarkRed, font=\boldmath] (q9) at (1.2, 0) {$q_9$};
      \node[vn, draw=green!60!black, text=green!60!black] (q10) at (2.1, 0) {$q_{10}$};

      % \draw[very thin, DarkRed] (q1) -- (X0);
      \draw[very thin, gray, densely dashed] (X1) -- (q1);
      \draw[very thin, DarkRed] (X1) -- (q2);
      \draw[->,  thick, gray, densely dashed] (X0) -- (q1);
      \draw[very thin, gray, densely dashed] (q3) -- (X2);
      \draw[very thin, DarkBlue] (q8) -- (X5);
      \draw[->,  thick, DarkRed] (X0) -- (q9);
      \draw[very thin, green!60!black] (q10) -- (X1);
      \draw[very thin, green!60!black] (q10) -- (dots);
      \draw[very thin, DarkRed] (q9) -- (X0);
      \draw[very thin, DarkRed] (q9) -- (X2);
    \end{tikzpicture}
    \caption{Symmetry breaking.}
    \label{fig:step3}
    \end{subfigure}
    \hfill
    % --- (e) Step 4 ---
    \begin{subfigure}[b]{0.19\linewidth}
    \centering
    \begin{tikzpicture}[
        scale=1.1,
        vn/.style={circle, draw=black, thick, minimum size=13pt, inner sep=0pt},
        cn/.style={rectangle, draw=black, thick, minimum size=13pt, inner sep=0pt}
    ]
      \node[cn, draw=DarkRed, text=DarkRed] (X0) at (0, 1) {$X_0$};
      \node[cn, draw=DarkRed, text=DarkRed] (X1) at (0.7, 1) {$X_1$};
      \node[cn, draw=DarkRed, text=DarkRed] (X2) at (1.4, 1) {$X_2$};
      \node[draw=none, text=black] (dots) at (1.9, 1) {$\dots$};
      \node[cn, draw=DarkBlue, text=DarkBlue] (X5) at (2.4, 1) {$X_5$};

      \node[vn, draw=gray, text=gray, densely dashed] (q1) at (0.3, 2) {$q_1$};
      \node[vn, draw=gray, text=gray, densely dashed] (q3) at (1.4, 2) {$q_3$};
      \node[vn, draw=DarkBlue, text=DarkBlue] (q8) at (2.4, 2) {$q_8$};

      \node[vn, draw=DarkRed, text=DarkRed, font=\boldmath] (q2) at (0.3, 0) {$q_2$};
      \node[vn, draw=DarkRed, text=DarkRed, font=\boldmath] (q9) at (1.2, 0) {$q_9$};
      \node[vn, draw=gray, text=gray] (q10) at (2.1, 0) {$q_{10}$};

      \draw[very thin, gray, densely dashed] (q1) -- (X0);
      \draw[->,  thick, gray, densely dashed] (X1) -- (q1);
      \draw[very thin, gray, densely dashed] (q3) -- (X2);
      \draw[very thin, DarkBlue] (q8) -- (X5);
      \draw[->,  thick, DarkRed] (X1) -- (q2);
      \draw[very thin, gray] (q10) -- (X1);
      \draw[very thin, gray] (q10) -- (dots);
      \draw[very thin, DarkRed] (q9) -- (X0);
      \draw[very thin, DarkRed] (q9) -- (X2);
    \end{tikzpicture}
    \caption{Final convergence.}
    \label{fig:step4}
    \end{subfigure}
    
    \caption{Example of randomness resolving a degenerate deadlock in a [[13,1,3]] surface code.
    We consider error events occurring on the data qubit set $\mathcal{I}_{e} = \{q_2, q_8, q_9\}$, which triggers the stabilizer signature $\mathcal{I}_{s} = \{X_0, X_1, X_2, X_5\}$.
    }
    \label{fig:lottery_workflow_combined}
\end{figure*}

\subsection{Potential of Randomness for Degeneracy}

Figure~\ref{fig:lottery_workflow_combined} illustrates how quantum degeneracy impacts BP and how randomness helps mitigate it.
\begin{itemize}
    \item Figure~\ref{fig:error_types}: $\{q_1, q_3\}$ and $\{q_2, q_9\}$ are degenerate errors, and both trigger the stabilizer $\{X_0, X_1, X_2\}$, while $\{q_8\}$ is a distinguishable error that uniquely triggers $\{X_5\}$.

    \item Figure~\ref{fig:step1}: The distinguishable error $\{q_8\}$ is easily decoded, resolving the syndrome $X_5$.
    However, BP struggles with the degenerate error candidates $\{q_1, q_3\}$ and $\{q_2, q_9\}$.
    As a result of this split belief~\cite{split_belief},  BP may output an ambiguous combination of these two error patterns, such as $\{q_1, q_9\}$, which leads to $X_0$ being 0, failing decoding.
    Such a deadlock has been observed in prior works~\cite{relaybp}.

    \item Figure~\ref{fig:step2}: By randomly perturbing the values of a small number of messages, a temporary error occurs at $q_{10}$ in green and redirects the message updates of stabilizer $X_1$ (green solid arrow).
    $X_1$ further biases $q_1$ towards the error-free state (gray dashed arrow), and biases $q_2$ towards the error state (red solid arrow).

    \item Figure~\ref{fig:step3}: As $q_1$ shifts toward the error-free state, the cancellation of parity checks at $X_0$ is resolved. This allows $X_0$ to further bias $q_1$ toward the error-free state (gray dashed arrow), and biases $q_9$ toward the error state (red solid arrow), successfully breaking the initial symmetry and amplifying the asymmetry between the degenerate errors.

    \item Figure~\ref{fig:step4}: Lacking support from syndromes, the temporary error at $q_{10}$ is suppressed ultimately. 
    Concurrently, $X_1$ further biases $q_1$ toward the error-free state (gray dashed arrow) and $q_2$ toward the error state (red solid arrow), leading the decoder to successfully converge.
    
\end{itemize}

\subsection{\name}
\label{sec:lotterybp}
To unleash the potential of randomness to mitigate quantum degeneracy, we propose \name.
We draw inspiration from prior works that flip the sign bit of the LLR of randomly selected VNs at each iteration to achieve better accuracy~\cite{BSFBP}.
However, it comes at the cost of significantly more decoding iterations, hurting the decoding latency~\cite{relaybp}.
% This method can improve BP accuracy with minimum cost.
% However, naively random selection of the VN imposes noticeable accuracy degradation to BP, and OSD fails to compensate for the accuracy drop, as random selection tends to mis-handle degeneracy errors.
As such, the problem to address is \textit{how to design an optimal lottery policy for random selection that maintains high accuracy with low hardware and iteration overhead}.

In the following, we explore the design space of the lottery policy to arrive at our final policy choice, and then detail its decoding algorithm.
% \todo{have you fixed the flow overall with preceding and subsequent paragraphs? the last paragraph flow is broken. also, we have the proposal lottery policy in design space of lottery policy now}

\subsubsection{Design Space of Lottery Policy}
\label{sec:lottery_design_space}
\begin{figure}[!t]
    \centering
    \begin{subfigure}[b]{0.48\columnwidth}
        \centering
        \includegraphics[width=\linewidth]{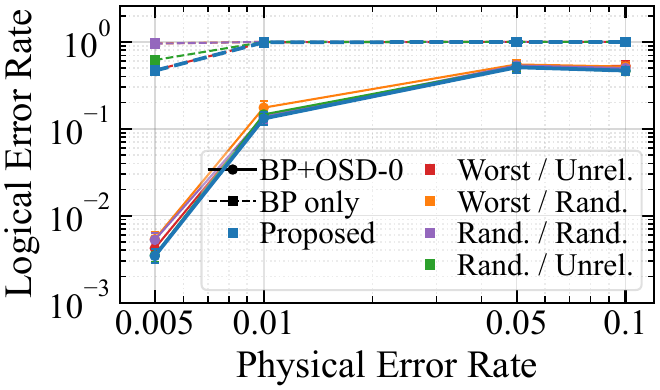}
        \caption{How to select lottery.}
        \label{fig:err_lottery_select}
    \end{subfigure}
    \begin{subfigure}[b]{0.48\columnwidth}
        \centering
        \includegraphics[width=\linewidth]{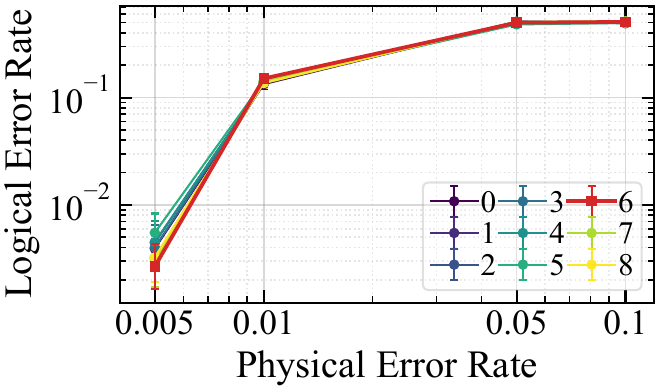}
        \caption{When to select lottery.}
        \label{fig:err_skip_iter}
    \end{subfigure}
    \caption{Lottery policies on surface code at $d=9$.
    }
    \label{fig:lottery_analysis}
\end{figure}

% In our example, split belief results in a symmetric fixed point, causing the decoder to fail to converge to a valid error pattern.
% As discussed in \cite{BSFBP}, distinguishable error patterns and their corresponding syndrome patterns can serve as auxiliary information to help resolve indistinguishable patterns through iterative message passing. 
% Perturbations introduced across iterations can give rise to new distinguishable error patterns, which provide guidance for decoding by interactive message passing, gradually leading to the dominance of one candidate error pattern.
% We analyze the design space of lottery policy (i.e., how to select lottery) and list five variants, where the three variants are adopted in prior works~\cite{BSFBP}. 
We analyze the design space of the lottery policy along two rules: which unsatisfied CN to target, and which of its neighboring VN to flip.
Each rule offers two choices, and a policy is a combination of one CN choice and one VN choice, yielding the four baseline policies compared in Table~\ref{tab:sf_policies}, with accuracy reported in Figure~\ref{fig:err_lottery_select}.
\begin{itemize}
    \item\textbf{CN selection:}
    \begin{itemize}
    \item\textbf{Worst CN:} 
    Select the unsatisfied CN connected to the largest number of VNs estimated to be in error.
    \item\textbf{Random CN:} 
    Select a random unsatisfied CN. 
    \end{itemize}
    \item\textbf{VN selection:}
    \begin{itemize}
    \item\textbf{Random VN:} 
    Flip the sign of a random neighboring VN of the selected CN  
    \item\textbf{Unreliable VN:} 
    Flip the sign of the neighboring VN of the selected CN with the minimum absolute LLR.
    \end{itemize}
    % \item \textbf{Global reliable:} 
    % Among all VNs with the most unsatisfied CNs, flip the sign of the one with the minimum absolute LLR. 
    % This policy builds an upper bound but is impractical for hardware implementation due to global search.
    % The decoder identifies variable nodes globally with the maximum number of unsatisfied check node connections and selects the one with the minimum LLR magnitude. 
    % This serves as a performance upper bound but is impractical for hardware due to the required global search.
    
    % \item \textbf{Global random:} 
    % Among all VNs with the most unsatisfied CNs, flip the sign of a random VN, ignoring LLR. 
    % This policy yields worse accuracy than global optimal, demonstrating the importance of reliability guidance.
    % that flipping high-impact nodes without reliability guidance can be detrimental.
    
    % \item \textbf{Local random:} 
    % For a randomly selected unsatisfied CN, flip the sign of a random neighboring VN.
    % Despite offering the lowest hardware complexity, the policy has low accuracy, due to lacking reliability guidance.
    
    % \item \textbf{Local reliable:} 
    % For a randomly selected unsatisfied CN, flip the sign of the neighboring VN with the minimum absolute LLR.
    % This policy lacks the connectivity-based prioritization, degrading accuracy over global optimal. (have way worse performance on proposed)
    
    \item \textbf{Proposed lottery policy:}
    % \todo{this sentence reads broken, also they are not updated. like I said, you need to read it thoroughly, some context before and after. I still see connectivity-based prioritization wording}
    Our proposed policy builds on top of the Random CN + Unreliable VN policy, with an additional local Worst CN selection.
    This additional selection solves the case where two VNs have identical LLRs (equally unreliable), and flip the VN with more unsatisfied CNs.
    % that considers only the CNs connected to the neighboring VNs of the selected CN.
    This simple update matches the accuracy of the Worst CN + Unreliable VN at low hardware cost.
    % , both with and without OSD.
    % Our proposed policy integrates these insights into a Random CN +  Unreliable VN policy. 
    % By combining worst CN with unreliability guidance within a local neighborhood, this policy achieves accuracy comparable to the global optimal at low hardware cost, both with and without OSD.
\end{itemize}

\begin{table}[t]
\centering
\caption{Comparison of sign-flipping policies.}
\label{tab:sf_policies}
\renewcommand{\arraystretch}{1.2}
\setlength{\tabcolsep}{4pt}
\footnotesize
\begin{tabular}{p{1cm}p{3cm}p{3cm}}
\toprule
 & \textbf{Unreliable VN} & \textbf{Random VN} \\
\midrule
\textbf{Worst CN}
& Best accuracy; global search prevents pipelined hardware, thus impractical to implement.
& Lower accuracy than ones with unreliable VN, highlighting the importance of unreliable VN. \\
\midrule
\textbf{Random CN}
& Simpler hardware implementation but lower accuracy due to the lack of global prioritization.
& Lowest hardware complexity but lowest decoding accuracy. \\
\bottomrule
\end{tabular}
\end{table}

% To find the most effective way to introduce this perturbation without compromising hardware efficiency, we explored the evolution of five different strategies, as compared in Figure \ref{fig:lottery_analysis}:

% The effectiveness of the \name lies in the combination of its two selection criteria. Unlike blind flipping, targeting a variable node with the highest connectivity to unsatisfied check nodes ensures that each perturbation has the maximum potential to reduce the syndrome weight. Simultaneously, using the minimum LLR magnitude as a tie-breaker ensures that the flip targets the least reliable qubit. This allows the decoder to introduce the necessary asymmetry into the message-passing updates without contradicting high-confidence beliefs, effectively steering the process away from symmetric fixed points at a minimal computational overhead.

% To visualize how these targeted perturbations successfully break a symmetric deadlock in practice, Figure~\ref{fig:lottery_workflow_combined} illustrates a step-by-step decoding workflow for a [[13,1,3]] surface code. When standard BP becomes trapped by the identical beliefs of degenerate pairs, \name injects a temporary auxiliary error based on the aforementioned rules. As shown in the figure, this perturbation asymmetrically reshapes the message updates, eliminates syndrome cancellations, and reactivates the necessary stabilizers. Ultimately, the auxiliary error naturally decays, guiding the decoder to converge onto a valid error configuration.

% sign flip
\begin{algorithm}[!t]
% \small
\caption{Proposed lottery policy for our \name, corresponding to line 13 in Algorithm~\ref{alg:decoding_nms_bp}.
All notations are identical to those in Algorithm~\ref{alg:decoding_nms_bp}.
\label{alg:decoding_lottory_bp}}
    $\mathcal{U}^{(i)} \gets \{c \mid \hat{s}_c^{(i-1)} \neq s_c\}$ \Comment*[r]{\CommentStyle{DarkRed}{Unsatisfied CN set}}
    $c^* \gets \textbf{RandomSelect}(\mathcal{U}^{(i)})$ \Comment*[r]{\CommentStyle{DarkRed}{Select random CN}}
    $\mathcal{V}_{max} \gets \underset{v \in \mathcal{V}(c^*)}{\arg\max} |\mathcal{C}(v) \cap \mathcal{U}^{(i)}|$ \Comment*[r]{\CommentStyle{DarkRed}{Max unsatisfied CN}}
    $v^* \gets \underset{v \in \mathcal{V}_{max}}{\arg\min} |\lambda^{(i)}_{v}|$ \Comment*[r]{\CommentStyle{DarkRed}{Min absolute LLR}}
    $\lambda^{(i)}_{v^*} \gets -\lambda^{(i)}_{v^*}$ \Comment*[r]{\CommentStyle{DarkRed}{Sign flip}}
\end{algorithm}

% \begin{figure*}[!t]
%     \centering
%     \begin{subfigure}[b]{0.3\textwidth}
%         \centering
%         \includegraphics[width=\linewidth]{fig/surf_dec_iter.pdf}
%         \caption{Surface code.}
%         \label{fig:surf_dec_iter}
%     \end{subfigure}
%     \begin{subfigure}[b]{0.3\textwidth}
%         \centering
%         \includegraphics[width=\linewidth]{fig/toric_dec_iter.pdf}
%         \caption{Toric code.}
%         \label{fig:toric_dec_iter}
%     \end{subfigure}
%     \begin{subfigure}[b]{0.3\textwidth}
%         \centering
%         \includegraphics[width=\linewidth]{fig/bb_dec_iter.pdf}
%         \caption{Bivariate Bicycle code.}
%         \label{fig:bb_dec_iter}
%     \end{subfigure}
%     \caption{Probabilistic distribution function of decoding iterations in BP, Relay-BP~\cite{relaybp}, and \name.}
%     \label{fig:dec_iter_analysis_bp_relay_lottery}
% \end{figure*}

\begin{figure}[!t]
    \centering
    \begin{subfigure}[b]{0.48\columnwidth}
        \centering
        \includegraphics[width=\linewidth]{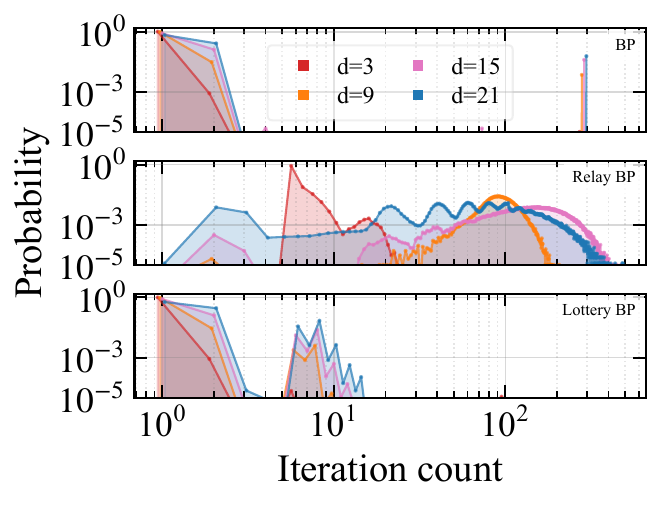}
        \caption{Surface code.}
        \label{fig:surf_dec_iter}
    \end{subfigure}
    % \begin{subfigure}[b]{0.48\columnwidth}
    %     \centering
    %     \includegraphics[width=\linewidth]{fig/toric_dec_iter.pdf}
    %     \caption{Toric code.}
    %     \label{fig:toric_dec_iter}
    % \end{subfigure}
    \begin{subfigure}[b]{0.48\columnwidth}
        \centering
        \includegraphics[width=\linewidth]{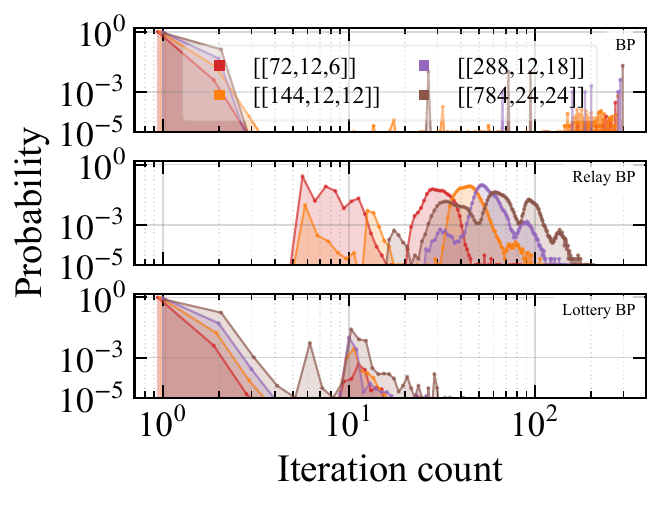}
        \caption{BB code.}
        \label{fig:bb_dec_iter}
    \end{subfigure}
    \caption{Distribution of decoding iterations at $p = 10^{-4}$.}
    \label{fig:dec_iter_analysis_bp_relay_lottery}
\end{figure}

% \subsubsection{Lottery Analysis}
% Figure~\ref{fig:lottery_analysis} compares the logical error rates of five perturbation variants to identify the optimal balance between decoding performance and hardware efficiency. 
% For standalone \name, the Local-Random, Global-Optimal, and Proposed \name strategies yield the best LER improvements. When integrating OSD post-processing, the top performers shift to Global-Optimal, Proposed \name, and Local-Reliable. 

% Across all configurations, Global-Optimal and our Proposed \name consistently deliver the lowest overall LER. The Proposed \name successfully matches this global performance upper bound while maintaining a strictly localized, hardware-friendly architecture.

% \todo{need to adjust the position of fig/table to match the order they appear in text}
\subsubsection{Detailed Proposed Lottery Algorithm}
% \todo{algorithm 3 should be before fig 5}
Algorithm~\ref{alg:decoding_lottory_bp} lists our proposed lottery policy to randomly select a VN and flip the sign of its LLR:

\begin{itemize}
    \item \textbf{Unsatisfied CN set (line~1):} 
    The unsatisfied CN set $\mathcal{U}^{(i)}$ is determined using the estimated syndrome from the previous iteration ($\hat{s}_c^{(i-1)}$). 
    This delay breaks the data dependency, enabling efficient pipelining (Section~\ref{sec:lottery_pipeline}).
    
    \item \textbf{Select random CN (line~2):} 
     Randomly select an unsatisfied CN $c^*$ from $\mathcal{U}^{(i)}$ to localize the perturbation region.
    
    \item \textbf{Max unsatisfied CN (line~3):} 
    Among all VNs connected to $c^*$, identify a subset $\mathcal{V}_{max}$ that connects to the most unsatisfied CNs.
    
    \item \textbf{Min absolute LLR (line~4):} 
    To ensure that the perturbation does not contradict strong reliability, select the target $v^* \in \mathcal{V}_{max}$ with the minimum absolute LLR.
    
    \item \textbf{Sign flip (line~5):} 
    Flip the sign of $v^*$'s LLR.
    % as perturbation.
\end{itemize}

Furthermore, we care about when to select lottery: whether we want to skip initial iterations for lottery; and if yes, how many iterations we should skip.
The results are given in Figure~\ref{fig:err_skip_iter}.
The rationale is that aggressively applying lottery to all iterations might perturb easy-to-decode syndromes, and reversely hurt the accuracy.
It is observed that skipping the first 6 iterations consistently wins at lower physical error rate ($0.5\times10^{-3}$), and we pick this for our \name.

\subsubsection{Decoding Iteration of \name}
The decoding iteration count for BP is defined as the number of message passing rounds to achieve convergence, i.e., when the estimated and measured syndromes perfectly match.
This metric directly dictates the decoding latency and energy consumption of the decoder hardware. 
While the worst-case latency is decided by code distance (Algorithm~\ref{alg:decoding_nms_bp}), lowering the probability of hitting this ceiling is crucial for overall system efficiency.
Figure~\ref{fig:dec_iter_analysis_bp_relay_lottery} compares the iteration distribution of BP, Relay BP (from IBM~\cite{relaybp}), and our \name on surface and Bivariate Bicycle (BB) code (from IBM~\cite{BB_code_paper}). 
For BP, significant decoding trials fail to converge, hitting the maximum iteration limit due to symmetric deadlocks.
Relay BP introduces a long-tail iteration distribution, orders of magnitude more than BP, similar to prior works introducing random sign flip~\cite{BSFBP}.
Both of them restart a brand new decoding process if the current process fails.
% In contrast, \name drastically reduces the proportion of these non-Relay BP relies on sequential restarts and weight modifications, it inherently produces a long-tail iteration distribution. 
\name mitigates hitting the maximum iteration and avoids restarting decoding, ensuring fast convergence within a single decoding process.
Hitting the maximum iteration less often also means higher accuracy (Section~\ref{sec:Evaluation}).

\section{Proposed \namearch Architecture}
\label{sec:Architecture}
% \todo{yanzhang}
We describe our decoding matrix construction and propose \namearch architecture that implement \name for low-latency local decoding and OSD for high-accuracy global decoding on surface code.
% Figure~\ref{fig:arch_overview} presents an overview of our \namearch and its pipelines.
% \name consists of three main blocks, V2C, C2V and Lottery, as depicted in Section~\ref{sec:arch_lottery_bp}, while OSD is described in Section~\ref{sec:arch_osd}.
% In the common case, decoding is performed using \name, which achieves fast convergence with low hardware cost. 
% Since OSD is invoked only when \name fails, a single OSD unit can be shared across multiple \name instances by reading the corresponding LLR values.

% x`
% The C2V processing pipeline includes (RC) read CN memory, (1C2V) 1-bit V2C conversion, (C2V) C2V conversion, (WU) write to unmatched syndrome register, (LLR) LLR update within VNUs (variable node units), (VN) variable node update within VNUs, (HSE) hard decision and syndrome estimation, (SC) syndrome comparison, (C2V) C2V conversion, (WV) write to VN memory, (VS) VN selector, and (CS) CN selector. The V2C processing pipeline includes (RV) read VN memory, (VS) Vn selector, (SF) sign flip, (CN) check node update within VNUs, (WC) write CN memory. The Bitonic sorter processing pipeline includes (RL) read LLR memory, (RLI) read LLR index memory, (BS) bitonic sorter, (WL) write LLR memory, (WLI) write LLR index memory. The LU decomposition and Forward processing pipeline includes (RLI) read LLR index memory, (LU) LU decomposition, (FW) forward substitution. The backward processing pipeline includes (BW) backward substitution. The block length of all modules has no relevance to execution time, except WV, CS, 1b C2V, RV and VS modules.

\subsection{Space-time matrix construction.}
% \todo{why need T here?}
To accommodate for $d$ measurement rounds, the Tanner graph is not the single-round parity-check matrix $H$ but the space-time matrix $H_{st}$, as shown in Equation~\ref{eq:h_st}. 

\begin{equation}
\label{eq:h_st}
H_{st} =
\begin{bmatrix}
H & 0 & 0 & \cdots & I & 0 & 0 & \cdots \\
0 & H & 0 & \cdots & I & I & 0 & \cdots \\
0 & 0 & H & \cdots & 0 & I & I & \cdots \\
\vdots & \vdots & \vdots & \ddots & \vdots & \vdots & \vdots & \ddots
\end{bmatrix}
\begin{matrix}
\scriptstyle\leftarrow\,\text{round }0 \\
\scriptstyle\leftarrow\,\text{round }1 \\
\scriptstyle\leftarrow\,\text{round }2 \\
\scriptstyle\ \ \vdots
\end{matrix}
\end{equation}

% \todo{how many reasons? whenever you update sth, dont just look they exact paragraphs, you need to think whether inconsistency are across the context}
Using our matrix construction, rather than the Detector Error Model (DEM) from Stim~\cite{stim}, is a critical trade-off between hardware cost and accuracy for two reasons.
% We avoid to using Stim detector-check matrix from the Detector Error Model (DEM) for three reasons:
% \todo{dont use ---, or --}
% \todo{technically, you can just say DEM is bad, you should say ours is way smaller (your current two points, can combine), deterministic to implement (regular), as well as introduce minimum errors}
First, a DEM check matrix from Stim carries one column per independent error source (every gate, reset, and measurement), so it has $13.8\times$ more columns than $H_{st}$ and carries over $25\times$ more nonzeros at $d{=}27$ (Table~\ref{tab:hst_size}), requiring $53$~Mb memory.
% at $d{=}27$ (Table~\ref{tab:hst_size}).
% \todo{still have ---, you need to use regular this word explicitly, then say this mean matrix can be generated on the fly. the current wording is not explicit}
Second, once the circuit is unrolled for processing, the matrix over rounds are no longer exact copies, meaning an irregular DEM matrix.
This irregular matrix can not be generated on the fly, requiring additional hardware for index mapping.
% Instead of being generated on the fly from the code distance d using a closed-form index mapping, the unrolled matrix must be stored explicitly.
% Lastly, collapsing the DEM into $H_{st}$ introduces only minimal decoding error \todo{need to add result here, it is one order of magnitude, so not minimal, consider removing this claim}, so the size and regularity gains come at negligible cost.

% \todo{still have T in table ii}
\begin{table}[t]
\centering
\footnotesize
\caption{Size of the $d$-round space-time decoding matrix (unrotated surface code): our generated $H_{st}$ (Equation~\ref{eq:h_st}) versus a circuit-level detector error model (DEM) extracted by Stim~\cite{stim}.
% Storage is the CSR sparse cost (nonzeros $\times\,\lceil\log_2(\text{cols})\rceil$ bit indices).
Storage is the memory required by the CSR representation (nonzeros $\times\,\lceil\log_2(\text{cols})\rceil$-bit column indices)
}
\label{tab:hst_size}
\setlength{\tabcolsep}{4pt}
\renewcommand{\arraystretch}{1.15}
\begin{tabular}{@{}ll rrrrr@{}}
\toprule
& & $d{=}3$ & $9$ & $15$ & $21$ & $27$ \\
\midrule
\multirow{2}{*}{Columns}
 & Stim & 423 & 23.9K & 125K & 361K & 788K \\
 & $H_{st}$ & 57 & 1.95K & 9.47K & 26.5K & 56.9K \\
\midrule
\multirow{2}{*}{Nonzeros}
 & Stim & 1.15K & 77.5K & 416K & 1.21M & 2.67M \\
 & $H_{st}$ & 96 & 3.74K & 18.5K & 52.1K & 112K \\
\midrule
\multirow{1}{*}{Storage}
 & Stim & 10\,Kb & 1.2\,Mb & 7.1\,Mb & 23\,Mb & 53\,Mb \\
\bottomrule
\end{tabular}
\end{table}

\subsection{\name}
\label{sec:arch_lottery_bp}
\begin{figure*}[!t]
    \centering
    \includegraphics[width=1\linewidth]{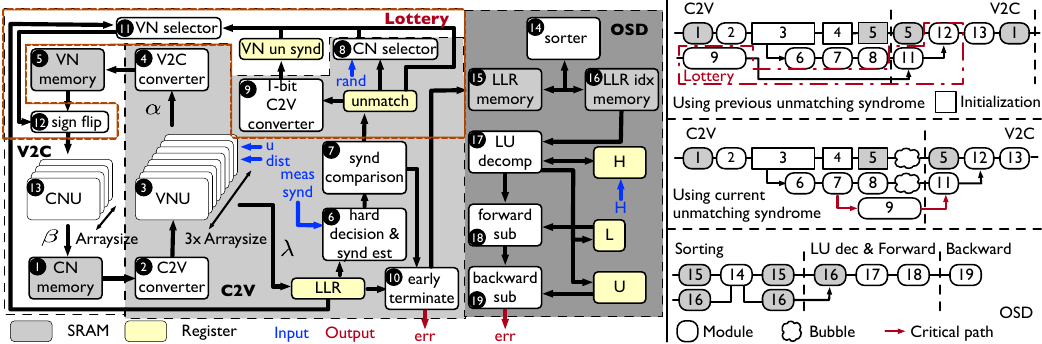}
    % \Description{The hardware implementation of the PolyQec architecture}
    \caption{Overview of \namearch. 
    On the right plane, vertical dashed lines separate each fully pipelined stages.}
    \label{fig:arch_overview}
\end{figure*}

In \namearch, one iteration of \name consists of two stages: V2C processing and C2V processing, with the workflow of \name following Algorithm~\ref{alg:decoding_nms_bp} and Algorithm~\ref{alg:decoding_lottory_bp}.
The \name pipeline is shown in Figure~\ref{fig:arch_overview}.

\minisection{Initialization.}
During the 0$^{th}$ iteration, the input channel LLR $\mu$ is initialized to \circled{3} VN update (VNU) to generate the V2C message $\alpha$.
Then \circled{4} V2C converter will convert $\alpha$ to the correct layout and write it to \circled{5} VN memory.

\minisection{V2C processing.}
After initialization, fully pipelined V2C processing starts with reading the V2C message $\alpha$ from \circled{5} VN memory.
After lottery is done via \circled{12} sign flip (Section~\ref{sec:lottery_pipeline}), the updated LLR $\lambda$ will be used to calculate the new C2V messages $\beta$ in \circled{13} CN update (CNU), which are further written back to the \circled{1} CN memory.
There is no layout conversion during V2C processing, since we align the memory layout in both VN and CN memory to CNU.

% begins with \circled{3} VN update (VNU), while other modules remain idle except \circled{4} V2C converter, \circled{5} VN memory, \circled{6} hard decision and syndrome estimation, \circled{7} syndrome comparison, and \circled{8}.
% The initialization iteration will produce the 
% This is only a partial pipeline of the C2V processing, as shown in the top row of the right plane of Figure~\ref{fig:arch_overview}.
% This behavior is consistent with the initialization phase described in Algorithm~\ref{alg:decoding_lottory_bp}. 
% The output of VNUs will go though the V2C converter, then the output of V2C converter will be written directly to the VN memory. 

% Once the initialization is done, the V2C processing starts immediately.
\minisection{C2V processing.}
After V2C processing, fully pipelined C2V processing will be launched. 
C2V processing starts with reading the C2V message $\beta$ from \circled{1} CN memory and convert $\beta$ to the correct layout in \circled{2} C2V converter.
Later, \circled{3} VNU calculates the new V2C message $\alpha$ and LLR $\lambda$.
The V2C message $\alpha$ will be sent to \circled{4} V2C converter, the same as in initialization.
The LLR $\lambda$ will go through \circled{6} hard decision and syndrome estimation.
If estimated and measured syndromes match in \circled{7} syndrome comparison, decoding succeeds in \circled{10} early termination; otherwise, continue to V2C processing.

The iterative V2C and C2V processing, if no successful decoding, will repeat until reaching the maximum iteration and invoke OSD.
Most hardware modules in \name are straightforward to design.
For example, VNU and CNU are just vector arrays of VN and CN, built from simple addition, subtraction, XOR, comparison, etc.
In the following, we will focus on describing memory layout and C2V/V2C converters to convert the message layout between VNU and CN/VN memories, as well as the lottery pipeline.

% As shown on the right-hand side of Fig.~\ref{fig:arch_overview}, V2C processing consists of reading messages from VN memory while simultaneously performing VN selection. 
% These two operations are synchronized such that their outputs arrive at the sign-flip module concurrently. 
% The selected messages are then sign-flipped, followed by check node updates (CNUs), and the results are written to CN memory.

% In contrast, C2V processing reads messages from CN memory while the 1-bit C2V conversion is performed in parallel. 
% The messages are then passed through the C2V converter, followed by variable node updates (VNUs), a subsequent V2C conversion, and finally written back to VN memory. 
% After the LLR computation on VNU stage, hard decision and syndrome estimation are performed, followed by syndrome comparison and CN selection.

% Through this alternating execution of V2C and C2V stages, \name decoding is realized. To enable \name functionality, additional components including the sign-flip module, VN selector, CN selector, and 1-bit C2V converter are integrated on top of the standard BP dataflow.

% In contrast, other components, such as VN update, CN update, and syndrome processing, are dominated by simple arithmetic operations (e.g., addition, subtraction, XOR, and comparison), which have regular dataflow and short critical paths, making them easier to pipeline and optimize.

% , as well as a few other optimizations.

\begin{table}[t]
\centering
\caption{The six VNs read for check node $i$, $d$-round space-time surface code,
where $m{=}d(d{-}1)$, $n{=}d^2{+}(d{-}1)^2$, $t{=}\lfloor i/m \rfloor$, and $j{=}i \bmod m$.
VN0 to VN3 are data qubits; VN4/VN5 are the syndrome connections between rounds.
Every pair is a fixed shift.
$\mathrm{VN} = \emptyset$ when:
For $H_X$, VN2: $j \bmod d = 0$, VN3: $j \bmod d = d{-}1$;
For $H_Z$, VN2: $\lfloor j/(d{-}1) \rfloor = 0$, VN3: $\lfloor j/(d{-}1) \rfloor = d{-}1$.}
\label{tab:cn_to_vn_indices}
\renewcommand{\arraystretch}{1.3}
\setlength{\tabcolsep}{2pt}
\scriptsize
\begin{tabular}{@{}p{0.15\linewidth}|p{0.30\linewidth}c|p{0.30\linewidth}c@{}}
\toprule
& \multicolumn{2}{c|}{\textbf{$X$ check}} & \multicolumn{2}{c}{\textbf{$Z$ check}} \\
\cmidrule(lr){2-3} \cmidrule(l){4-5}
\textbf{Pair} & \textbf{first VN} & \textbf{shift} & \textbf{first VN} & \textbf{shift} \\
\midrule
VN0$\rightarrow$VN1
  & $j + tn$ & $+d$
  & $j + \lfloor j/(d-1) \rfloor + tn$ & $+1$ \\
VN2$\rightarrow$VN3
  & $d^2 + j - 1 - \lfloor j/d \rfloor + tn$ & $+1$
  & $d^2 + j - d + 1 + tn$ & $+(d{-}1)$ \\
VN4$\rightarrow$VN5 
  & $Tn + i - m$ & $+m$
  & $Tn + i - m$ & $+m$ \\
\bottomrule
\end{tabular}
\end{table}

\subsubsection{Memory Layout and Conversion}
\begin{figure}[!t]
    \centering
    \includegraphics[width=1\linewidth]{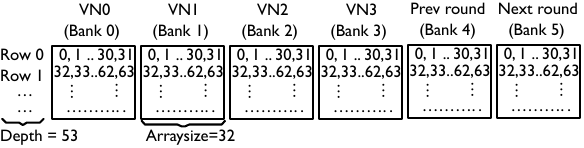}
    \caption{CN memory layout with array size of 32 and $d = 15$.}
    \label{fig:memory_bank}
\end{figure}

All V2C and C2V messages are stored in \circled{5} VN and \circled{1} CN memory.
We adopt CN layout for both the memories, which means CNU directly reads the memory and work on the data without any conversion, allowing fully pipelining in V2C processing.
We exemplify this layout in Figure~\ref{fig:memory_bank}.
We provide six memory banks, and each bank stores the messages for one of the six VNs connecting a CN (Connection between rounds will be seen as VNs). 
Though we mark them as VN0$\sim$5, the actual VN indices differ across CN and are determined by the parity-check matrix.
In CN layout, each row of all banks stores the same group of adjacent CNs, and the row size equals the array size (the number of CNs in the CNU).
For example, with an array size of 32, each bank will store 32 different CNs at each row, e.g., from CN0 to CN127 at row 0.
At each clock cycle, all fix memory banks will be read simultaneously.
The output of VN memory will be used by \circled{13} CNU to generate the new C2V message $\beta$, while the output of the CN memory will be converted by \circled{2} C2V converter.
% Due to VN and CN memories adopt the same CN layout, V2C processing is fully pipelined.
In surface codes, since one CN connects fix VNs but one VN connects only two CNs, the array size of VNU is twice that of CNU, to ensure fully pipelining in C2V processing.
% Therefore, reading four banks of data enforces twice the array size of CNU in VNU.
For CNs connecting 2 or 3 VNs (syndrome qubits at the corner or edge), we store these messages as zero.
Table~\ref{tab:cn_to_vn_indices} summarizes the mapping from CN to VN indices for memory layout conversion, which is calculated from the parity-check matrix.
% , facilitating layout conversion.
% For these CNs, there are missing connections with VNs, causing only 2 or 3 valid messages, and we will store these messages as zero.
% and mask them out in CNU and VNU.

% Given CN memory as an example, it is partitioned into 4 banks, each single value in a bank repesent the message $\beta^{(i)}_{v\leftarrow c}$ of corresponding variable node $v$ which connected to the current check node $c$ at iteration i, for each iteration on the algorithm, the full VN and CN memory will be updated.
% If the $ARRAYSIZE$ = 16 and distance = 11, then the full reading from CN memory at $2^{nd}$ bank and $1^{st}$ address for $3^{rd}$ iteration will be 
% \[[\beta^{(3)}_{v^{12}\leftarrow c^{1}}, \beta^{(3)}_{v^{13}\leftarrow c^{2}},....\beta^{(3)}_{v^{26}\leftarrow c^{15}}, \beta^{(3)}_{v^{27}\leftarrow c^{16}}]\].

% Understanding the memory layout, we now describe the design layout conversion.
Compared to other modules, layout conversion for V2C and C2V messages in V2C and C2V converters is critical, due to supporting different connectivity patterns, as defined by the parity-check matrix.
In classical BP decoding for LDPC codes, these connectivity patterns have to be stored in the memory as a sparse matrix~\cite{7419905}, whose cost increases quadratically with the code distance.
Fortunately, the parity-check matrices of surface codes are regular, which facilitates converting the message layout on the fly instead of storing the parity-check matrix. 
\circled{2} C2V converter converts the C2V message from CN layout to VN layout, which groups the messages needed by each VN together.
% Similar to the CN layout exemplified in Figure~\ref{fig:memory_bank}, VN layout means the C2V message can be fed to VNs without any transformation.
% corresponding to the same VN are aligned (e.g., the first and second entries correspond to the same VN index, and similarly for the third and fourth entries). 
On the other hand, the V2C converter performs the inverse conversion, from VN layout to CN layout.
These two converters allow full pipelined C2V processing.
% This approach enables both CN and VN updates to be executed directly on aligned messages without requiring lookup tables or storage of the full parity-check matrix.
% Table~\ref{tab:cn_to_vn_indices} summarizes the transformation in the C2V converter, which is calculated from the parity-check matrices.
% \todo{table III might be too much, we can delete it or leave it to appendix, if we need more space}
Using Table~\ref{tab:cn_to_vn_indices}, C2V messages are routed to correct VNs.
If a VN index is mapped to VNU for processing at the current cycle, the message will be consumed;
otherwise, buffered until the VN index is mapped.
% using multiplexing and synchronize the messages with delay.
% For example, in the surface code X check, the VN0 and VN1 messages need a $d$-cycle delay via a shift register for synchronization.
% For simplicity, we skip the implementation.
% For example, in the surface code X check, the C2V converter can be efficiently implemented using a ping-pong buffer and deterministic shifts. 
% The VN1 is shifted forward by $d$ to aligning it with the VN0 stream. Similarly, the fourth VN stream is shifted forward by one position to align with the third VN stream. 
% As a result, messages corresponding to the same VN are co-located and can be processed in parallel. The V2C converter reverses this process by shifting the second VN stream backward by $d$ and the fourth VN stream backward by one position, restoring the original CN layout. 
% The similar approach can be applied to the surface code $H_Z$, as well as both $H_X$ and $H_Z$ of the toric code, following the relationships summarized in Table~\ref{tab:cn_to_vn_indices}.

\subsubsection{Optimized Lottery Pipeline}
\label{sec:lottery_pipeline}
After \circled{7} syndrome comparison, the mismatch between the measured syndromes from the previous iteration (line 1 of Algorithm~\ref{alg:decoding_lottory_bp}) and estimated syndromes is forwarded to \circled{8} CN selector. 
CN selector selects a CN based on a random number by counting mismatches.
Specifically, for $N$ mismatches and a random value $r \in [0,1)$, the selected CN index is $\lfloor r \cdot N \rfloor$. 
% For example, with 50 mismatches and $r=0.70$, the $35^{\text{th}}$ mismatched CN is selected. 
% We use the unmatched syndrome from the previous iteration (line 1 of Algorithm~\ref{alg:decoding_lottory_bp}) to ensure fully pipelined V2C processing.
The 1-bit C2V converter will convert the unmatches from CN layout to VN layout.
\circled{11} VN selector identifies all VNs connected to the selected CN and determines which VN is associated with the largest number of unsatisfied CNs using results from \circled{9} 1-bit C2V converter. 
In the case of a tie, LLRs are compared to select the VN with minimum absolute LLR.
% most likely \todo{"most likely -> minimum absolute LLR or Least reliable"?, I'm not sure whether "most likely" would be misleading} VN. 
A mask is then generated to isolate the selected VN and forwarded with its LLR to \circled{12} sign flip module.
Using the unmatched syndrome from the previous iteration is critical to ensure full pipelining.
If we use the unmatched syndrome from the current iteration, we would have the pipeline shown in the middle of the right plane in Figure~\ref{fig:arch_overview}.
Here, the 1-bit C2V converter stays on the critical path and causes pipeline stalls.

\subsection{OSD}
\label{sec:arch_osd}
OSD runs three stages sequentially (Figure~\ref{fig:arch_overview}): (1) sorting, (2) LU decomposition with forward substitution, and (3) backward substitution.
The reliability LLRs from BP are read from \circled{15} and sorted in \circled{14}, with their indices tracked by \circled{16} LLR index memory; \circled{17} LU decomposition then consumes the sorted LLRs, fully pipelined with \circled{18} forward substitution, and \circled{19} backward substitution completes OSD.

\begin{figure}[!t]
    \centering
    \includegraphics[width=1\linewidth]{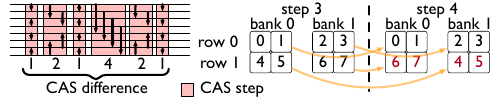}
    \caption{Resolving read bank conflict of LLR memory.}
    \label{fig:llr_swap}
\end{figure}

We implement a bitonic sorter.
Existing sorters apply maximally parallel comparison-and-swap (CAS) units to the full input~\cite{chen2015bitonic, salamat2021nascent}, yielding low utilization at low code distances; \namearch instead scales to 65536 LLRs ($d\le27$) with a small CAS array.
In Figure~\ref{fig:llr_swap}, each step (red block) issues multiple CAS, whose CAS difference is the gap between its two input indices.
LLRs and indices each use two banks holding array-size adjacent entries per row, all read per cycle.
Once the CAS difference reaches twice the array size, both CAS inputs fall in one bank and stall on a read bank conflict.
Predicting this from the next step's difference, we swap the write-back address: step 4 has difference $4=2\times2$, so step 3 writes 0/1 and 4/5 to banks 0 and 1 for parallel reads.

LU decomposition and both substitutions reduce to row/column swaps, XOR, and sum.
The lower triangular matrix \(L\) is generated column by column; generating column $i$ freezes row $i$, which then feeds forward substitution immediately, fully pipelining the two stages.
Backward substitution cannot, since rows of the upper triangular matrix \(U\) are still modified after their column is generated.
We store \(H\), \(L\), \(U\) as dense boolean matrices, since their number of 1s grows during decomposition.

\section{Proposed \namesim Simulator}
\label{sec:Software}
% \todo{yanzhang}
\begin{figure}[!t]
    \centering
    \includegraphics[width=1\linewidth]{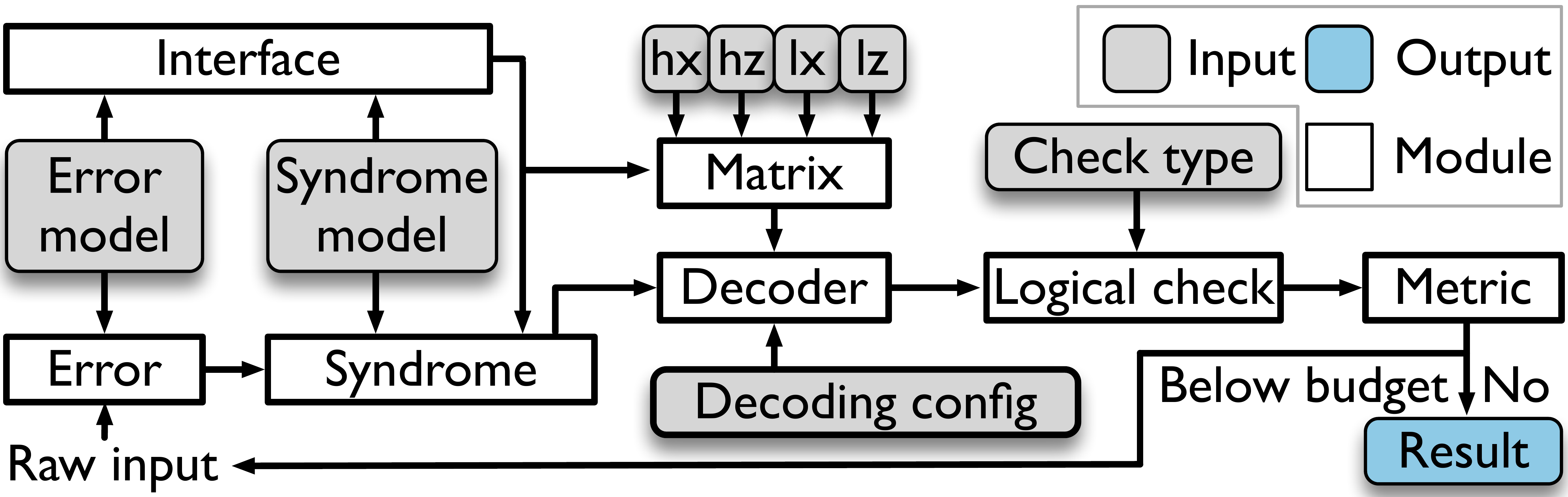}
    \caption{Overview of \namesim.}
    \label{fig:syndrilla_framework}
\end{figure}

% \subsection{Overview}
Figure~\ref{fig:syndrilla_framework} outlines \namesim, our QEC decoding simulator.
The workflow of \namesim is similar to other QEC decoding simulators~\cite{bposdcpp, afs_paper, astrea, das2022lilliput, nvidia_qec_decoder, deltakit, swamit_decoder_bench}, including error and syndrome generation, followed by decoder and logical check.
What distinguishes \namesim is the integrated metrics for fair evaluation and cross-platform execution for accelerated simulation.
\namesim also interacts with Stim~\cite{stim} to support broader error models.

\minisection{Integrated Metrics.}
\namesim integrates multifaceted metrics to evaluate decoding accuracy, runtime, and efficiency, which is missing in prior works.
% Rather than considering individual metrics in isolation, we categorize them into two groups that reflect the key objectives of QEC decoder design.
Accuracy metrics target end-to-end algorithm evaluation, including logical error rate and detailed ratio of data/syndrome qubit errors are identified and/or corrected.
Runtime metrics compare simulation speed, include runtime per simulation, batch, sample and iteration, allowing users to understand the bottleneck of the simulator.
Efficiency metrics hint hardware efficiency, including iteration count and distribution, converge failure and success rate, and invoke rate for each decoder (Section~\ref{sec:invoke_rate_global}), allowing users to understand the algorithm bottlenecks.
Together, these multi-faceted metrics allows to evaluate decoders fairly. 

\begin{figure}[!t]
    \centering
    \begin{subfigure}[b]{0.48\columnwidth}
        \centering
        \includegraphics[width=\linewidth]{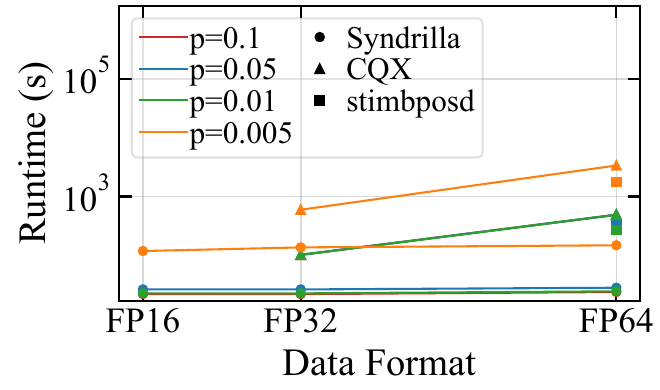}
        \caption{Runtime vs data format.}
        \label{fig:runtime_format}
    \end{subfigure}
    \hfill
    \begin{subfigure}[b]{0.48\columnwidth}
        \centering
        \includegraphics[width=\linewidth]{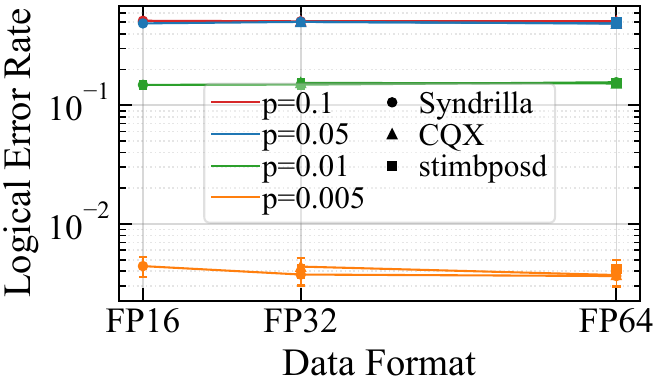}
        \caption{Accuracy vs data format.}
        \label{fig:acc_format}
    \end{subfigure}
    \begin{subfigure}[b]{0.48\columnwidth}
        \centering
        \includegraphics[width=\linewidth]{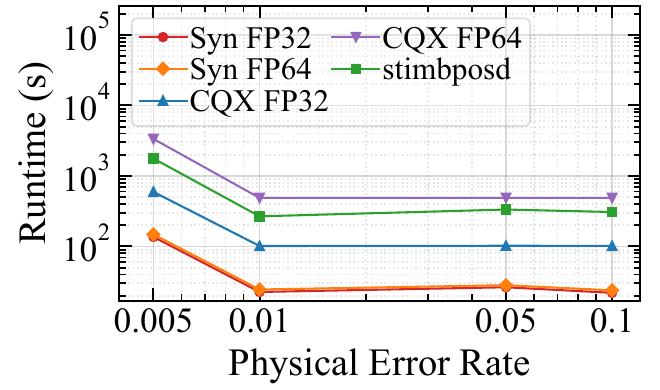}
        \caption{Runtime vs platform.}
        \label{fig:runtime_distance}
    \end{subfigure}
    \hfill
    \begin{subfigure}[b]{0.48\columnwidth}
        \centering
        \includegraphics[width=\linewidth]{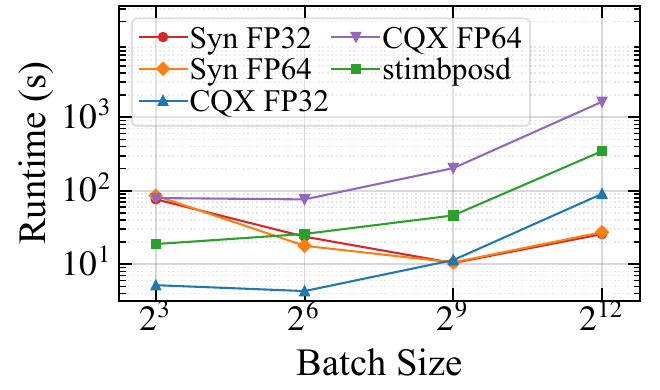}
        \caption{Runtime vs batch size.}
        \label{fig:runtime_batch}
    \end{subfigure}

    \caption{Simulating surface code at $d=9$. 
    CUDAQX (CQX), StimBPOSD and \namesim are all running on Intel Core i5-13450HX and NVIDIA GeForce RTX 4090. 
    The runtime represents the total simulation time for each configuration.
    (a), (b) and (c) have a batch size of 4,000 and targer error number of 100 for BP+OSD.
    stimbposd only support FP64 and CUDAQX only support FP64 and FP32.
    (d) uses $p=0.05$ and target error $=100$.}
    \label{fig:runtime_analysis}
\end{figure}

\minisection{Accelerated Simulation.}
% \todo{yanzhang}
\namesim is built on PyTorch to benefit from cross-platform execution, unlike other simulators, which target CPUs~\cite{bposdcpp, deltakit, swamit_decoder_bench} or GPUs from a single vendor~\cite{nvidia_qec_decoder}.
Every decoder in \namesim additionally has a CUDA implementation, so acceleration is available across the entire decoder suite.
Figure~\ref{fig:runtime_analysis} shows the simulation accuracy and speedup of \namesim.
\namesim supports a broader range of data formats, FP16, FP32, and FP64 for simulation (Figure~\ref{fig:runtime_format} and \ref{fig:acc_format}).
% \todo{c/d has weird dashed lines, i dont think necessary, since you already use different colors}
% \todo{have you defined CQX the first time it is mentioned in the caption?}
\namesim on GPU is more than an order of magnitude faster than close-sourced CUDAQX~\cite{nvidia_qec_decoder} and StimBPOSD~\cite{bposdcpp} across physical error rates and batch sizes (Figure~\ref{fig:runtime_distance} and \ref{fig:runtime_batch}).
This speedup becomes increasingly critical at lower error rates, which demand trillions of simulation shots.
\namesim on CPU runs slower than StimBPOSD because each batch must synchronize to its slowest sample.
Simulation exhibits distinct sweet spots in batch size.

% \todo{zhenyou, what is the runtime here, per sample? per batch? per full simulation? i guess per batch, double check, update the caption if wrong; squeeze the height of the figs, check whether d use 0.05 physical error rate, check c batch size}

% Based on Fig.~\ref{fig:runtime_distance}, the runtime of Syndrilla for BP+OSD is approximately one order up to two order of magnitude faster than the C++ implementation~\cite{bposdcpp} as the physical error rates increase, This improvement is primarily due to Syndrilla’s ability to exploit parallelism across large batch sizes.

% Similarly, as shown in Fig.~\ref{fig:runtime_batch}, Syndrilla achieves a speedup of approximately one to two orders of magnitude over the C++ implementation as the batch size increases. Under the same settings, \name consistently outperforms standard BP in runtime, as it exhibits a higher convergence rate and lower logical error rate, reducing the need for additional iterations or fallback decoding.

% Finally, the choice of data format does not impact decoding accuracy or runtime performance.

\section{Experimental Setup}
\label{sec:Implementation}

\minisection{Code Configuration.}
% \todo{zhenyou, yunhao}
% \dummytext{1}
% Following prior works~\cite{afs_paper, Micro_blossom, Vegapunk}, we adopt the standard circuit-level noise model, which accounts for both data qubit errors and measurement errors, for our evaluations.
Following prior works~\cite{afs_paper, Micro_blossom, Vegapunk}, we adopt a standard circuit-level noise model parameterized by physical error rate $p$, extracted from Stim~\cite{stim}. 
Under this model, all single-qubit operations, two-qubit gates, idle periods, and measurements fail with probability $p$. 
Exploiting the CSS code symmetry, our simulations independently track X-type errors.
% ~\cite{dephasing_channel_paper, dephasing_channel_paper_2}; 
% Specifically, data qubits are subject to independent bit-flip (X) and phase-flip (Z) errors with a physical error rate $p$. 
% Measurement errors are then independently introduced to syndromes with an identical probability $p$ at each measurement round.
% We focus on X errors via Monte Carlo simulation, due to the symmetricity of X and Z errors under our noise model.
% \todo{zhenyou, double check measurement error description is accurate.}
% which i also used in prior works~\cite{afs_paper, Micro_blossom, Vegapunk}.
% where each physical qubit experiences a Pauli-$X$ or Pauli-$Z$ error with probability $p$ respectively. 
% We evaluate decoder performance using Monte Carlo experiments, under a noise model composed of independent bit-flip and phase-flip channels. This model can be viewed as two independent pure \emph{dephasing noise model} \cite{dephasing_channel_paper} \cite{dephasing_channel_paper_2}, where each physical qubit experiences a Pauli-$X$ or Pauli-$Z$ error with probability $p$ respectively. 
% Due to the symmetry of the noise model, $X$-type and $Z$-type experiments are equivalent, allowing the performance observed in $Z$-type memory experiments to be directly extrapolated to $X$-type errors. Therefore, we evaluate only $Z$-type experiments in this study.
We simulate two code families: the surface code (for code distances up to $d=27$) and BB codes \cite{BB_code_paper} (up to $d=24$). The physical error rate $p \in 5 \times 10^{-3}\sim10^{-1}$. Note that simulations for the $d=27$ surface code were omitted due to GPU out-of-memory.
To ensure low variance, we stop each simulation when a sufficient number of failed trials is reached, e.g., 100.
% , in order to obtain statistically reliable estimates of the logical error rate (LER). 
A trial is marked as failed if the logical measurement differs from the initial state.
% \todo{zhenyou, 100 or 1000 for target error? also}

\minisection{Decoding Simulator.}
% \todo{zhenyou, yunhao}
% \dummytext{2}
% As BP-based baselines, we employ the conventional normalized min-sum BP decoder and the BP decoder with OSD-0 post-processing \cite{bposdcpp} decoder.
We simulate normalized min-sum BP~\cite{norm_min_sum_paper} (Algorithm~\ref{alg:decoding_nms_bp}) and OSD-0~\cite{bposdcpp} (Algorithm~\ref{alg:decoding_osd0}) as the local and global decoders, using \namesim with FP32 data.
We also analyze Relay BP~\cite{relaybp} and BP-SF~\cite{BP_SF} without OSD.
For non-BP decoders, we consider low-complexity heuristic-based UF~\cite{perlin2023qldpc} and near-optimal MWPM~\cite{Sparse_blossom}.
% \todo{cite the code for UF and MWPM}
% better than worst case
% \cite{afs_paper}
% Promatch: Extending the Reach of Real-Time Quantum Error Correction with Adaptive Predecoding
% \dw{error model: whether syndrome error is presented?}
% double check multi-round measurement and decoding

\minisection{Hardware Implementation.}
We evaluate the hardware area and power of AFS~\cite{afs_paper} and Vegapunk~\cite{Vegapunk} based on their estimation models.
Note that AFS only includes the memory cost.
We synthesize and place-and-route Micro Blossom~\cite{Micro_blossom} based on their open-source RTL.
These baselines' hardware results will scale with the code distance.
For \namearch, we implement all modules in Verilog, synthesize and place-and-route them.
We choose the array size to be 256 for both \name and OSD, and the memory is fixed to support up to $d=27$, meaning our hardware cost is constant across code distances.
% \todo{yanzhang, add array size here}
BP and Relay BP are adapted from our \name.
All synthesis and place-and-route are done using FreePDK45 45nm at 400MHz.
SRAM area and power are obtained from CACTI7~\cite{cacti7}.
We adapt all implementations to larger code distances using an open-source tool for design space exploration~\cite{A-Graph}, by scaling up the required array width, memory size, etc.

\section{Evaluation}
\label{sec:Evaluation}

% \todo{add these insight to where they fit}
% \insight[ins:lottery]{lottery help with degeneracy errors in BP.}
% \insight[ins:vote]{majority vote before decoding help suppress measurement errors in BP.}
% \insight[ins:vote_before_dec]{vote-before-decoding allows more latency margin for decoding, and alleviate backlog problem.}
% \insight[ins:local_dec]{design better local decoders significantly benefits scaling up the systems..}

% we focus on evaluating X errors.

% Different types of error will be using the same physical error for testing.\cite{Micro_blossom}\cite{afs_paper}\cite{}
\subsection{Algorithm Analysis}
% \todo{zhenyou, yunhao}

\subsubsection{Logical Error Rate}

\begin{figure}[!t]
    \centering
    % ：Surface code
    \begin{subfigure}[b]{0.48\linewidth}
        \centering
        \includegraphics[width=\linewidth]{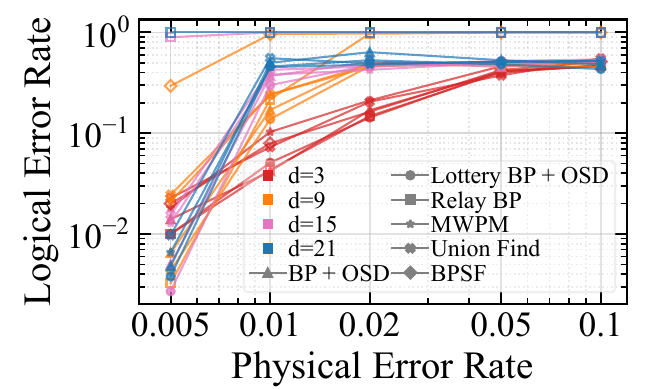}
        \caption{Surface code with OSD.}
        \label{fig:surf_bp}
    \end{subfigure}
    \hfill
    \begin{subfigure}[b]{0.48\linewidth}
        \centering
        \includegraphics[width=\linewidth]{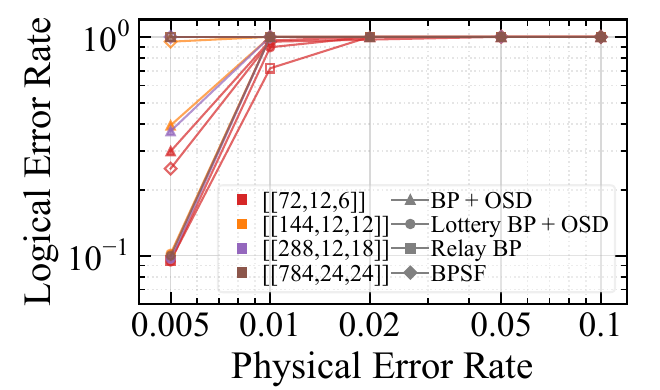}
        \caption{BB code. MWPM fails.}
        \label{fig:bb_osd}
    \end{subfigure}
    % \begin{subfigure}[b]{0.48\linewidth}
    %     \centering
    %     \includegraphics[width=\linewidth]{fig/SURFACE_stim_DEM_advanced.pdf}
    %     \caption{Surface code with OSD.}
    %     \label{fig:surf_osd}
    % \end{subfigure}

    % ：Toric code
    % \begin{subfigure}[b]{0.48\linewidth}
    %     \centering
    %     \includegraphics[width=\linewidth]{fig/toric_lottery_BP.pdf}
    %     \caption{Toric code without OSD.}
    %     \label{fig:toric_bp}
    % \end{subfigure}
    % \hfill
    % \begin{subfigure}[b]{0.48\linewidth}
    %     \centering
    %     \includegraphics[width=\linewidth]{fig/toric_lottery_BP_OSD.pdf}
    %     \caption{Toric code with OSD.}
    %     \label{fig:toric_osd}
    % \end{subfigure}

    % ：Bivariate Bicycle code
    % \begin{subfigure}[b]{0.48\linewidth}
    %     \centering
    %     \includegraphics[width=\linewidth]{fig/BB_stim_DEM_pure_bp.pdf}
    %     \caption{BB code without OSD.}
    %     \label{fig:bb_bp}
    % \end{subfigure}
    % \hfill
    % \begin{subfigure}[b]{0.48\linewidth}
    %     \centering
    %     \includegraphics[width=\linewidth]{fig/BB_stim_DEM_advanced.pdf}
    %     \caption{BB code with OSD.}
    %     \label{fig:bb_osd}
    % \end{subfigure}
    
    \caption{Logical error rate in Stim DEM.
    % , with and without OSD.
    % \dw{zhenyou, reduce the margin of all figs you have produced}
    }
    \label{fig:err_analysis_dem}
\end{figure}

% \begin{figure*}[!t]
%     \centering
%     \begin{subfigure}[b]{0.3\textwidth}
%         \centering
%         \includegraphics[width=\linewidth]{fig/surf_uf_mwpf_relay_lottery.pdf}
%         \caption{Surface code.}
%         \label{fig:surf_uf_mwpf_relay_lottery}
%     \end{subfigure}
%     \begin{subfigure}[b]{0.3\textwidth}
%         \centering
%         \includegraphics[width=\linewidth]{fig/toric_uf_mwpf_relay_lottery.pdf}
%         \caption{Toric code.}
%         \label{fig:toric_uf_mwpf_relay_lottery}
%     \end{subfigure}
%     \begin{subfigure}[b]{0.3\textwidth}
%         \centering
%         \includegraphics[width=\linewidth]{fig/bb_uf_mwpf_relay_lottery.pdf}
%         \caption{BB code.
%         MWPF fails on QLDPC code.}
%         \label{fig:bb_uf_mwpf_relay_lottery}
%     \end{subfigure}
%     \caption{Logical error rate of UF, MWPF, Relay BP, and \namefull, simulated with measurement errors.}
%     \label{fig:err_analysis_uf_mwpf_relay_lottery}
% \end{figure*}

\begin{figure}[!t]
    \centering
    \begin{subfigure}[b]{0.48\columnwidth}
        \centering
        \includegraphics[width=\linewidth]{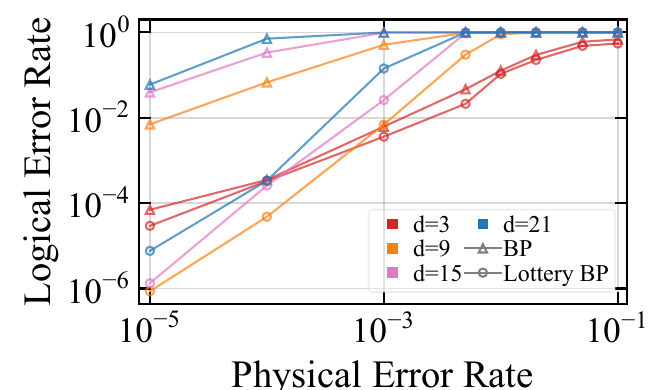}
        \caption{Surface code without OSD.}
        \label{fig:surf_uf_mwpf_relay_lottery}
    \end{subfigure}
    \hfill
    \begin{subfigure}[b]{0.48\linewidth}
        \centering
        \includegraphics[width=\linewidth]{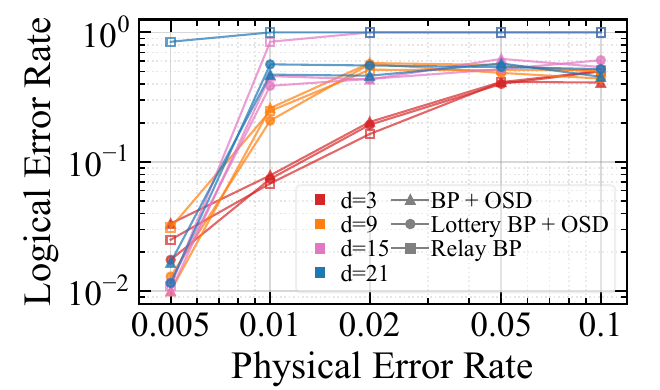}
        \caption{Surface code with OSD.}
        \label{fig:surf_osd}
    \end{subfigure}
    \begin{subfigure}[b]{0.48\linewidth}
        \centering
        \includegraphics[width=\linewidth]{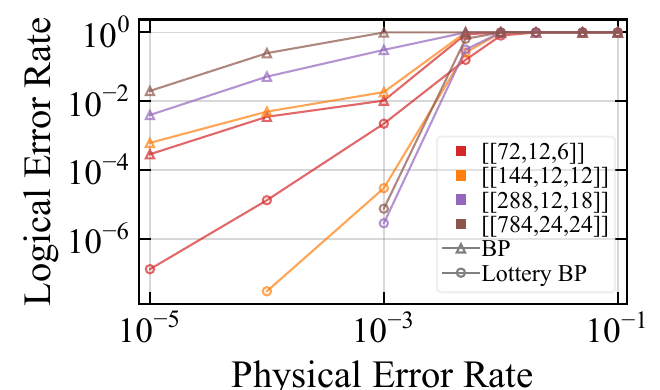}
        \caption{BB code without OSD.}
        \label{fig:bb_bp}
    \end{subfigure}
    \hfill
    \begin{subfigure}[b]{0.48\columnwidth}
        \centering
        \includegraphics[width=\linewidth]{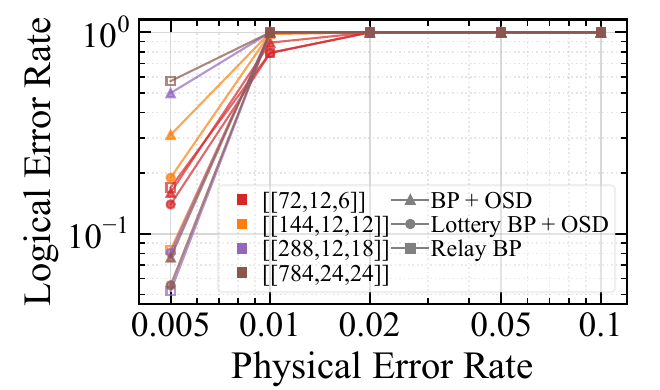}
        \caption{BB code with OSD.}
        \label{fig:bb_uf_mwpf_relay_lottery}
    \end{subfigure}
    \caption{Logical error rate in space-time matrix.}
    \label{fig:err_analysis_st}
\end{figure}

Our logical error rate evaluations are conducted in a relatively high physical error rate regime, with maximum 300 BP iterations.
Results are shown in Figure~\ref{fig:err_analysis_dem} and Figure~\ref{fig:err_analysis_st}.
Across both surface and BB codes, \name consistently achieves significantly lower error rates than standard BP, a performance advantage that becomes increasingly pronounced as the physical error rate decreases.
When augmented with OSD, \name on surface codes outperforms BP-SF, achieving a competitive accuracy comparable to both standard BP+OSD and pure Relay BP, while simultaneously benefiting from significantly reduced decoding latency (Figure~\ref{fig:latency_analysis}). 
Similarly, for BB codes, \namefull matches the accuracy of pure Relay BP and outperforms both standard BP+OSD and BP-SF.
It is worth noting that across all evaluated codes, our 3D space-time matrix formulation uses a phenomenological graph to decode circuit-level noise. 
While this method dramatically reduces decoding complexity, it incurs an accuracy penalty, yielding logical error rates approximately one order of magnitude higher than those obtained using the exact Stim DEM graphs (Figure~\ref{fig:err_analysis_dem} vs Figure~\ref{fig:err_analysis_st}).

\subsubsection{Invoke Rate of Global Decoder}
\label{sec:invoke_rate_global}
% Figure~\ref{fig:err_analysis_uf_mwpf_relay_lottery} further compares the decoding accuracy with measurement error.
% For surface code, \namefull ranks the second in decoding accuracy, next to the best MWPM.
% For BB code, \namefull is the best, since MWPM does not work for QLDPC code.
% This demonstrates the effectiveness of syndrome vote to address measurement errors.
\begin{figure}[!t]
\centering
    \begin{subfigure}[b]{0.48\linewidth}
        \includegraphics[width=\linewidth,clip]{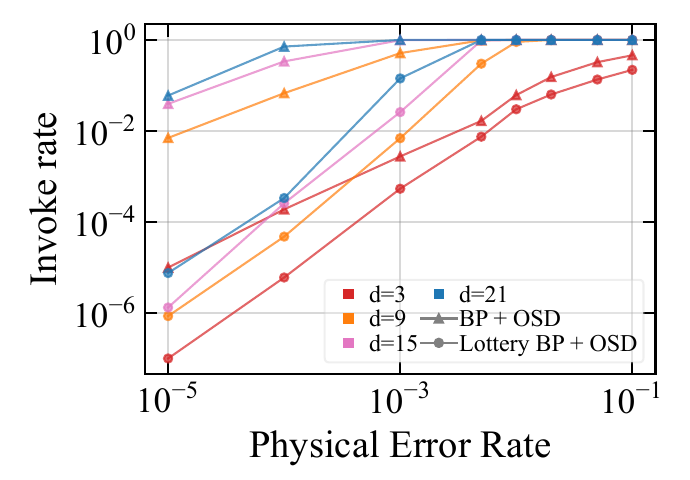}
        \caption{Surface code.}
        \label{fig:surf_not_conv}
    \end{subfigure}
    \hfill
    % ... toric panel (still commented out) ...
    \begin{subfigure}[b]{0.48\linewidth}
        \includegraphics[width=\linewidth,clip]{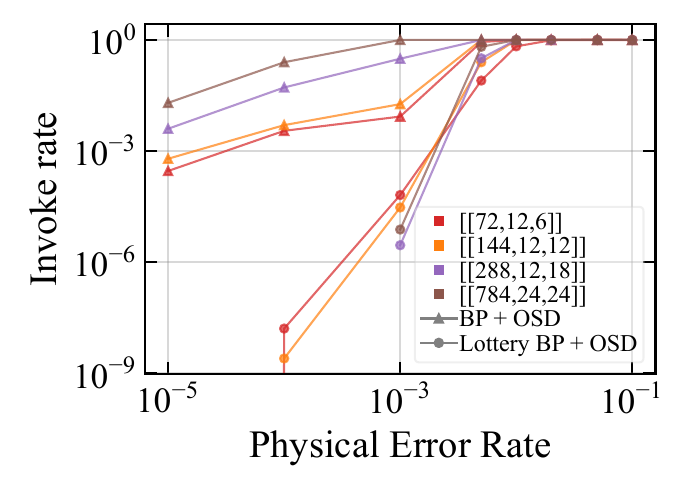}
        \caption{BB code.}
        \label{fig:bb_not_conv}
    \end{subfigure}
\caption{Invoke rate of OSD.}
\label{fig:invoke_rate_analysis_uf_mwpf_relay_lottery}
\end{figure}

In local-global decoder, the costly global OSD is invoke when the local BP fails to converge, penalizing the decoding latency and efficiency.
% While OSD is accurate, its high hardware cost (sorting and matrix operations) penalizes the decoding latency and efficiency.
To ensure scalable QEC, one has to invoke the global decoder less.
Figure~\ref{fig:invoke_rate_analysis_uf_mwpf_relay_lottery} compares the invoke rate of OSD on all codes.
Compared to standard BP, \name substantially reduces the OSD invoke rate for both surface and BB codes, with this reduction being even more pronounced for BB codes. 
% Compared to \name, BP exhibits substantially higher invoke rates for surface codes, and slightly higher rates for BB code.
% In contrast, \name successfully forces convergence on a massive fraction of these locally trapped errors. 
% Consequently, \namefull dramatically suppresses the required OSD invocations across all evaluated code distances and topologies. 
\name achieves up to 4 orders of magnitude lower invoke rate for surface code.
% For certain toric codes (e.g., $d=3$), \name almost never invokes OSD.
% This massive reduction in global decoder workload does not incur any degradation in error correction capability. 
% As previously established, the convergence failure rate of \namefull is comparable to that of standard BP+OSD. 
Even for BB code where BP performs quite well, \name invokes up to 6 orders of magnitude less.
By resolving most errors in the local decoder and invoking the global decoder less, \namefull enables more scalable decoding, thus QEC. 
% \todo{is here not finished yet? or I'm wrong}

% \begin{figure}[!t]
%     \centering
    
%     \begin{subfigure}[b]{0.48\linewidth}
%         \includegraphics[width=\linewidth]{fig/surf_not_conv.pdf}
%         \caption{Surface (Not conv.)}
%         \label{fig:surf_not_conv}
%     \end{subfigure}
%     \hfill
%     \begin{subfigure}[b]{0.48\linewidth}
%         \includegraphics[width=\linewidth]{fig/surf_conv_failure.pdf}
%         \caption{Surface (Conv. fail)}
%         \label{fig:surf_conv_fail}
%     \end{subfigure}
    
%     \vspace{0.5em}
    
%     \begin{subfigure}[b]{0.48\linewidth}
%         \includegraphics[width=\linewidth]{fig/toric_not_conv.pdf}
%         \caption{Toric (No-conv.)}
%         \label{fig:toric_not_conv}
%     \end{subfigure}
%     \hfill
%     \begin{subfigure}[b]{0.48\linewidth}
%         \includegraphics[width=\linewidth]{fig/toric_conv_failure.pdf}
%         \caption{Toric (Conv. fail)}
%         \label{fig:toric_conv_fail}
%     \end{subfigure}
    
%     \vspace{0.5em}

%     \begin{subfigure}[b]{0.48\linewidth}
%         \includegraphics[width=\linewidth]{fig/BB_not_conv.pdf}
%         \caption{BB code (Not conv.)}
%         \label{fig:bb_not_conv}
%     \end{subfigure}
%     \hfill
%     \begin{subfigure}[b]{0.48\linewidth}
%         \includegraphics[width=\linewidth]{fig/BB_conv_failure.pdf}
%         \caption{BB code (Conv. fail)}
%         \label{fig:bb_conv_fail}
%     \end{subfigure}
    
%     \caption{Breakdown of decoding failures. 
%     The not converge and converge fail rates means no valid .}
%     \label{fig:invoke_rate_analysis_uf_mwpf_relay_lottery}
% \end{figure}

\subsubsection{Quantization Analysis}
% this weird shape is because too close to fig 16.
% \begin{wrapfigure}{R}{1.165\linewidth}
\begin{wrapfigure}{R}{0.48\linewidth}
\begin{center}
    \includegraphics[width=\linewidth,clip]{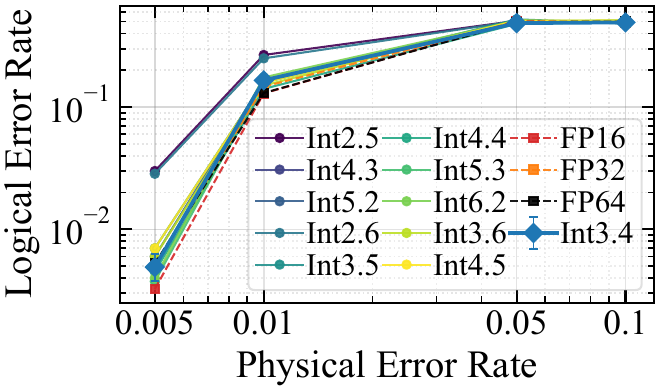}
\end{center}
\caption{Quantization on surface code at $d=9$.}
\label{fig:quant_analysis}
\end{wrapfigure}

Unlike using floating-point in simulation, we need to quantize the soft messages to fixed-point format for efficient hardware implementation.
Figure~\ref{fig:quant_analysis} compares the decoding accuracy of \namefull, with $x.y$ for $x$-bit integer and $y$-bit fraction.
Different floating-point formats behave closely, as validated in Figure~\ref{fig:acc_format}.
7-bit fixed-point suffers from severe truncation noise, leading to noticeable accuracy drop. 
Increasing to 8 bits, Int3.4 perfectly match the floating-point baselines. 
We choose Int3.4 in our hardware implementation, offering optimal accuracy-cost trade-off.
% Consequently, 8-bit fixed-point arithmetic provides the optimal hardware-accuracy trade-off for low-power hardware design.

% \begin{figure}[!t]
%     \centering
%     \includegraphics[width=0.48\linewidth]{fig/err_quantization.pdf}
%     \label{fig:err_quantization}
%     \caption{Fixed-point on surface code at $d=9$.
%     }
%     \label{fig:quant_analysis}
% \end{figure}

\subsection{Architecture Analysis}
% \todo{yanzhang}

Since hardware baselines claim a wide range of running frequency, from 43MHz implemented in Micro Blossom~\cite{Micro_blossom} to 4GHz assumed in AFS~\cite{afs_paper}, we assume an identical running frequency at 1GHz for fair comparison in the hardware evaluation.
We evaluate hardware on surface code for X check with $p= 10^{-5}$, following prior works~\cite{afs_paper, Micro_blossom, Vegapunk, astrea}.
For baseline decoder architectures, including AFS~\cite{afs_paper}, Micro Blossom~\cite{Micro_blossom}, and Vegapunk~\cite{Vegapunk}, the required computational resources are highly dependent on the code distance $d$. 
For example, AFS and Micro Blossom operate on a 3D space-time graph, where the number of vertices and edges (thus area, power, and latency) scale with both the code distance and physical error rate.
On the contrary, our implementation of \namearch supports up to $d=27$ with fixed area and power, but different latency.

% baseline: 
% \name on archx is done, all done, decoupled BP and OSD implementation
% AFS (UF): memory is done on archx, logic is not
% vegapunk (customized)
% Micro Blossom (MWPF)
% relay-BP

% area/latency/power study is done across error rate/code distance
% they should have breakdown.

% For architecture analysis part, Archx\cite{A-Graph} is used on analysis QolyQec and comparing with other baselines. By splitting the full hardware design into small pieces. Each part can be scale simply. Although baselines like Vegapunk\cite{Vegapunk}, Micro Blossom\cite{Micro_blossom} only show the analysis on FPGA, and for AFS\cite{afs_paper} the paper only shows
% memory analysis. However, Micro Blossom has open-source verilog code, and Vegapunk can be implemented based on the given method. Archx\cite{A-Graph} helps analysis the Micro Blossom and Vegapunk.

\subsubsection{Latency}
\begin{figure}[!t]
    \centering
    \begin{subfigure}[b]{0.48\columnwidth}
        \centering
        \includegraphics[width=\linewidth]{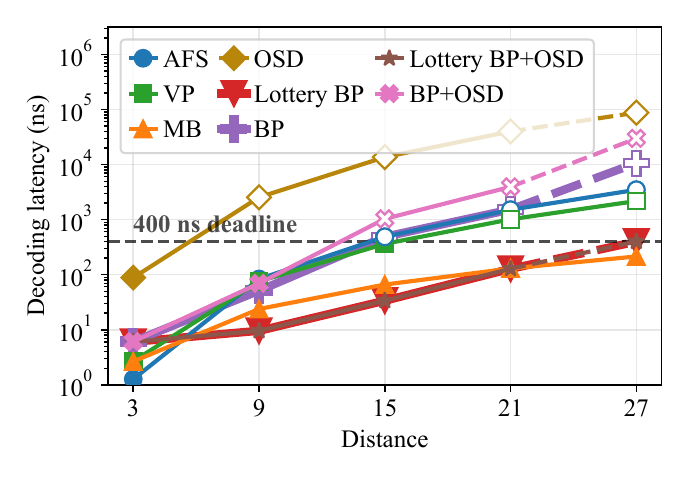}
        \caption{Average latency vs distance.}
        \label{fig:latency_vs_distance}
    \end{subfigure}
    \begin{subfigure}[b]{0.48\columnwidth}
        \centering
        \includegraphics[width=\linewidth]{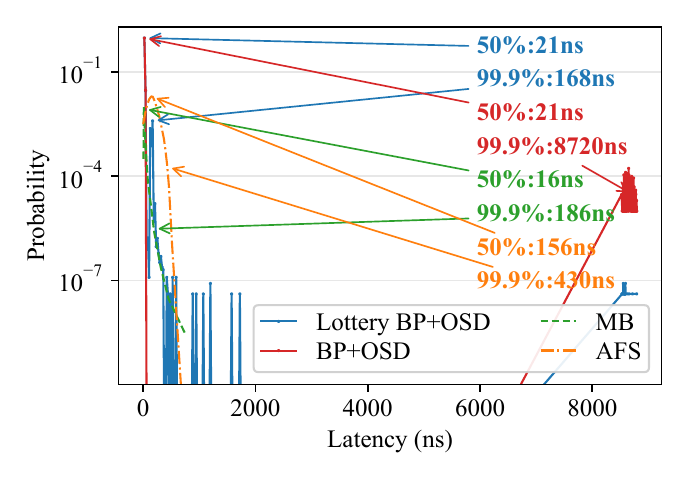}
        \caption{Latency distribution.}
        \label{fig:latency_distribution}
    \end{subfigure}
    \caption{Latency. 
    VP and MB are short for Vegapunk and Micro Blossom (CPU time excluded).
    (b) use $d=9$. 
    }
    \label{fig:latency_analysis}
\end{figure}

Figure~\ref{fig:latency_analysis} compares the latency of different decoder hardware.
Figure~\ref{fig:latency_vs_distance} shows how latency changes across code distances.
OSD shows the highest latency among all decoders, due to its costly implementation for sorting and matrix operations.
Due to the low invoke rate, OSD's contribution to total latency of \namefull is also negligible, e.g., the latencies of \name and \namefull are almost identical.
\namefull also shows lower latency than BP+OSD, as \name converges faster (Figure~\ref{fig:dec_iter_analysis_bp_relay_lottery}).
Overall, our \namefull always exhibits best latency across most evaluated distances, while other decoders based on 3D decoding graph often violate the 400ns decoding latency at large code distances~\cite{surface_cycle}.
Although Micro-Blossom reports a comparable decoding latency, it employs a heterogeneous CPU–accelerator architecture in which part of the decoding is executed on an ARM Cortex-A72 processor. Consequently, its reported latency is not achieved by the hardware accelerator alone, unlike \namefull.
Figure~\ref{fig:latency_distribution} shows the distribution of decoding latency.
Compared to AFS and Micro Blossom, the latency of \namefull and BP+OSD are more centered at a low range.
\namefull has the 50th and 99.9th percentile decoding latencies as 21 ns and 168 ns, showing a low variance.
For \namefull, BP+OSD, Micro Blossom and AFS, the 99.99th percentile decoding latencies are 540.00 ns, 11728.00 ns, 321.17 ns, and 574.07 ns.
\namefull invokes OSD less than BP+OSD to avoid the long tail latency.

\subsubsection{Area}
% \todo{legend coloring not aligned in a/b, also double check related numbers}
Figure~\ref{fig:area_vs_distance} compares the area across code distances.
Note that AFS reports the SRAM area, while all BP variants have a fixed SRAM size supporting up to $d=27$.
Vegapunk does not have an SRAM area, since its memory is implemented with register file.
All BP variants occupy similar area, with \name having $26.71mm^2$.
Micro Blossom does not require the largest hardware area because much of the MWPM computation is performed by the CPU (ARM Cortex-A72), with the hardware accelerator implementing only part of the decoding algorithm.
AFS has the smallest area among all, since UF is cheaper to implement and, more important, logic area is excluded~\cite{afs_paper}.
Unsurprisingly, the OSD decoder has the second largest area, $202.05\times$ larger than \name at $d=27$.
This indicates the high cost of a global decoder and raises the need for a low invoke rate.
At $d=27$, \name has $1.02\times$ larger area than BP, a negligible overhead.
Importantly, \name exhibits $0.049\times$, $1083.83\times$, and $1.27\times$ lower area than AFS, Vegapunk, and Micro Blossom.
% \todo{add numbers.}
% is able to have a way lower area cost in term of local decoder. 
% Both Vegapunk and Micro Blossom has a global part which could be treated as the global decoder as OSD. 
% As Micro Blossom, Vegapunk is grow rapidly as the distance increase, while all three types of BP will be slower. 
% And \name + OSD is reconfigurable to any code distance with CN size less than the actual number of values that 1 memory bank can be stores.

\subsubsection{Power}
\begin{figure}[!t]
    \centering
    \begin{subfigure}[b]{0.48\columnwidth}
        \centering
        \includegraphics[width=\linewidth]{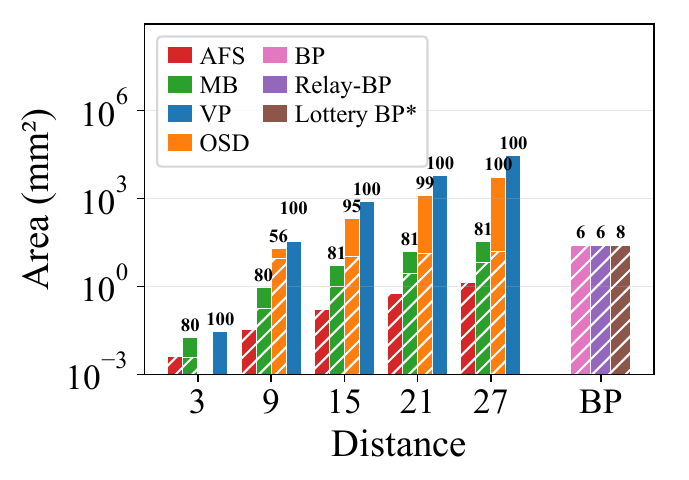}
        \caption{Area vs distance.
        For each stack bar, the top (numbers are its ratio) and bottom (hatched) are logic and SRAM area.
        }
        \label{fig:area_vs_distance}
    \end{subfigure}
    \begin{subfigure}[b]{0.48\columnwidth}
        \centering
        \includegraphics[width=\linewidth]{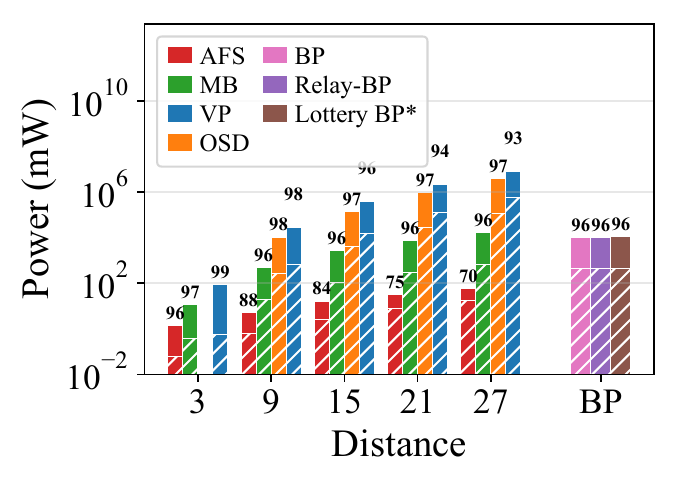}
        \caption{Power vs distance.
        For each stack bar, the top (numbers are its ratio) and bottom (hatched) are dynamic and leakage power.
        }
        \label{fig:power_vs_distance}
    \end{subfigure}
    \caption{Hardware for space-time matrix. 
    BP supports $d\le27$.
    }
    \label{fig:area_power_analysis}
\end{figure}

Figure~\ref{fig:power_vs_distance} compares the power across code distances.
Again, BP variants show similar power consumption.
Note here we only report the AFS based on its memory size, underestimating the total power.
% Though this underestimates its total power consumption, AFS's power is still larger than \name at $d=32$.
At $d=27$, \name has the third lowest power ($10909.84$mW) compared with AFS, Vegapunk, and Micro Blossom, showing $0.0052\times$, $740.49\times$, and $1.60\times$ improvements.
Though OSD has high power ($4035037.59$mW), with a low invoke rate ($1.82 \times 10^{-5}$), its contribution to the total power is negligible.
For the full \namearch with OSD invoke rate counted, the power is $34935.93$mW, $0.0016\times$, $231.24\times$, and $0.50\times$ better than AFS, Vegapunk, and Micro Blossom.
% Here, we use the invoke rate at $d=13$ and $p=10^{-3}$ for power estimation, since $d=13$ is the closest distance we have simulated to $d=32$.
% We consider this invoke rate conservative, since the invoke rate shows a dropping trend with large distances, as given in Figure~\ref{fig:invoke_rate_analysis_uf_mwpf_relay_lottery}.
The mW-level power renders \name promising for local decoding in cryogenic domain, with OSD as a global decoder in room temprature.
% As illustrated in Figure~8(b), all BP-based designs (\name, Relay BP, and conventional BP) are capable of supporting surface codes up to a distance of $d=32$. In contrast, all other decoder designs exhibit significantly higher power consumption. 

% Even when compared with AFS at $d=32$—which is evaluated using an SRAM only modering still shows higher power consumption than all BP-based designs. Moreover, when accounting for the OSD stage, \name maintains an extremely low invocation rate on the order of $10^{-6}$. For example, at $d=9$, the invocation rate is $1.534 \times 10^{-6}$, implying that only one OSD invocation is required for approximately every 652,174 BP decoders.

% Accordingly, the average power consumption of a BP+OSD system can be expressed as the power of BP plus the product of the OSD invocation rate and the power of OSD. Under this model, the average power of BP-based designs with OSD remains significantly lower than that of all other decoder architectures.

\subsubsection{Scalability}
% system level scaling
% where to put the local and globle decoder? project room temp CMOS results to cryogenic CMOS, and then calculate the additional power needed. model the bandwidth needed to ensure latency
\begin{wrapfigure}{R}{0.48\linewidth}
\begin{center}
    \includegraphics[width=\linewidth,clip]{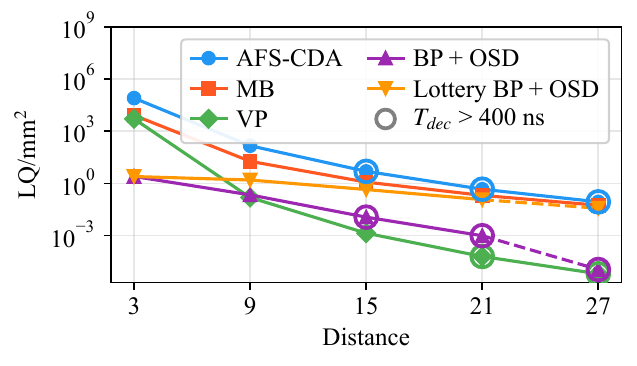}
\end{center}
\caption{Scalability study. 
}
\label{fig:lq_per_mm2}
\end{wrapfigure}
To support millions of qubits, decoding has to be massively parallel within a tight latency margin due to syndrome generation~\cite{surface_cycle}.
An optimal decoder is expected to decode as many logical qubits as possible, with minimum area and time.
We quantify this advantage using a novel metric, \textit{decoding efficiency}, defined in Equation~\ref{eq:dec_eff}.
\begin{equation}
\label{eq:dec_eff}
    \textit{Decoding efficiency} = \frac{\textit{Num of decoding}}{\textit{Area}_\textit{decoder}} \cdot \frac{T_{\textit{margin}}}{T_{\textit{decoder}}}
\end{equation}
Here, $T_{\textit{margin}}$ is the latency margin to complete the decoding, e.g., 400ns for superconducting qubits in conventional decoders~\cite{surface_cycle}, and $T_{\textit{decoder}}$ means average decoding latency.
% With our syndrome vote to handle measurement errors, \namefull decode only once per $d$ measurement rounds, meaning $T_{\textit{margin}}=400d$~ns for \namefull.
% be able to  We want to further understand how scalable \namearch is, and 
% As quantum systems scale to larger code distances ($d$), global decoders like MWPM or OSD suffer from severe latency growth. 
% This poor scalability creates a critical bottleneck: if the decoding rate ($r_{proc}$) falls behind the syndrome generation rate ($r_{gen}$), a catastrophic \textbf{decoding backlog} emerges. Such backlogs induce exponential latency overheads in quantum algorithms, ultimately destroying any quantum advantage.
The prolonged latency margin mitigates the backlog problem~\cite{terhal2015quantum, holmes2020nisq+}, unlocking scalable quantum error decoding.
% effectively maintains the crucial condition $r_{proc} > r_{gen}$ under practical operational regimes. Consequently, our architecture scales efficiently  mitigates the risk of accumulating decoding backlogs in large-scale quantum systems.

In this evaluation, one OSD is shared by multiple \name (Figure~\ref{fig:design_space_overview} right), scaled by the invoke rate.
Figure~\ref{fig:lq_per_mm2} compares the decoding efficiency of different decoders.
Our \namefull achieves the third-best decoding efficiency, up to 4 orders of magnitude better than BP+OSD due to lower OSD invoke rate.
The two higher-ranked designs are not directly comparable: AFS reports only memory cost, while MB offloads much of the decoding to CPU.
% , with the hardware accelerator implementing only part of the decoder.
% \todo{yanzhang, fig 20 change ''No MS'' to ''No syndrome vote''}
% Without syndrome vote, \namefull's decoding efficiency drops by $d$, but still outperforms others.
Note that \namefull uses a fixed memory size to support $d\le27$, while memories of other decoders are distance-optimized, causing AFS-CDA (conjointed decoder architecture for resource saving~\cite{afs_paper}) and Vegapunk to win at small distances.

\FloatBarrier
\section{Related Works}
\label{sec:Discussion}

\minisection{Decoding Algorithm.}
% \todo{zhenyou, yunhao}
To mitigate the accuracy drop, BSFBP~\cite{BSFBP} introduces branched decoding with sign flipping at the cost of prohibitively more iteration.
Our \name breaks degeneracy without latency penalties.
Another prominent approach is MWPM~\cite{MWPM_1, MWPM_2, MWPM_3}. 
Recent software implementations~\cite{Sparse_blossom, Fusion_blossom} and hardware accelerators~\cite{Micro_blossom} have achieved ultra-low latency with high accuracy.
However, MWPM is fundamentally restricted to topological codes where errors trigger at most two syndrome defects. 
\name overcomes this topology restriction while maintaining efficiency and scalability with minor accuracy drops.

\minisection{Decoder Architecture.}
% \todo{yanzhang}
Hardware-efficient decoder architectures must meet strict latency and scalability requirements in fault-tolerant quantum computing, while balancing decoding accuracy, latency, and hardware cost.
MWPM-based designs, such as Astrea~\cite{astrea} and Micro-Blossom~\cite{Micro_blossom}, achieve high accuracy but suffer from poor scalability due to high complexity.
% due to high computational complexity and resource overhead. 
In contrast, Union-Find decoders, like AFS~\cite{afs_paper}, provide lower complexity and better scalability through efficient pipelined architectures, though with some loss in accuracy. 

%Multi-DR decoder architecture~\cite{10.1145/3620665.3640388}, parallel decoders in 4K and 300K temperature. Heterogeneity in decoding at scale~\cite{10.1145/3620665.3640388}.4K for intra-DR patches, 300K for inter-DR patches.

\minisection{Decoding Simulator.}
% \todo{yanzhang}
% mention NVIDIA work\cite{nvidia_qec_decoder}, bposd\cite{bposdcpp}, stim\cite{stim}, etc.
% decoder bench from swamit\cite{swamit_decoder_bench};
% deltakit\cite{deltakit};
Stim~\cite{stim} is widely used for simulating stabilizer circuit and generate syndrome under realistic noise models. but not for decoding errors.
% \todo{add a reference for bposd+stim repo}
BPOSD~\cite{stimbposd} has been extended with Stim for integrated CPU simulation.
Deltakit and decoder-bench~\cite{swamit_decoder_bench} further provide modular, standardized frameworks for comparing QEC decoders on top of Stim.
CUDA-Q~\cite{cudaq, nvidia_qec_decoder} provides a NVIDIA GPU-centered platform for constructing QEC codes, generating syndrome circuits, and evaluating decoding workflows at scale, with support for real quantum hardware real-time testing.

\section{Conclusion}
\label{sec:Conclusion}
Targeting scalable quantum error decoding, we propose \name to improve the local decoding accuracy while reducing global decoder invoke rate by up to 4 orders for surface code, so as to reduce the average decoding latency. 
% We propose syndrome vote before \name to compress the multiple rounds of measured syndromes to one, allowing more decoding latency and mitigating the backlog problem.
% \todo{check numbers, read carefully around, not just one sentence}
The implemented architecture \namearch can support up to 4 orders of magnitude more logical qubits than BPOSD~\cite{stimbposd} under the same decoding budget.
Our decoding simulator \namesim offers more than 1 order of magnitude speedup.
% which can significantly lower the logical rate about 2 order which will also increase as the distance increases, lower the invoke rate of global decoder OSD for about 2 order with a few small hardware components, without having impacts on preformance when combined with OSD.
% To hardware implementation on the Lottery BP + OSD, we introduce the PolyQec, which has much lower area cost and higher scalability compared to all other baselines. In addition, For measurement error problem which most 2D space decoders are facing, we propose the majority vote strategy to keep the logical error similar to the result without measurement error and to give longer decoding time margin which is the original decoding time margin X d.

% for acknowledgement
% \begin{acks}
% \end{acks}

%%%%%%%%% -- BIB STYLE AND FILE -- %%%%%%%%
\bibliographystyle{IEEEtranS}
\bibliography{ref}

\end{document}